\documentclass[12pt]{modifiediopart}

\usepackage{graphicx,epsfig,array,amssymb}

\renewcommand{\theequation}{\thesection.\arabic{equation}}

\newcommand{\MC}{m_\mathrm{c}}
\newcommand{\RI}{r_\mathrm{0}}
\newcommand{\MR}{m_\mathrm{r}}
\newcommand{\MG}{m_\mathrm{g}}
\newcommand{\RC}{r_\mathrm{c}}
\newcommand{\TC}{t_\mathrm{c}}
\newcommand{\CS}{c_{\mathrm s}}
\newcommand{\RMAX}{r_\mathrm{max}}

\renewcommand{\theequation}{\thesection.\arabic{equation}}

\begin{document}

\title{On the dynamics of relativistic multi-layer spherical shell systems}

\author{Merse E.\,G\'asp\'ar and  Istv\'an R\'acz}

\address{\small RMKI,
\small  H-1121 Budapest, Konkoly Thege Mikl\'os \'ut 29-33.
\small Hungary}

\ead{merse@rmki.kfki.hu, iracz@rmki.kfki.hu}

\begin{abstract}
The relativistic time evolution of multi-layer spherically symmetric shell systems---consisting of
infinitely thin shells separated by vacuum regions---is examined.
Whenever two shells collide the evolution is continued with the assumption that the collision is totally transparent.
The time evolution of various multi-layer shell systems---comprised by large number of shells
thereby mimicking the behavior of a thick shell making it possible to study the formation of acoustic singularities---is analyzed numerically
and compared in certain cases to the corresponding Newtonian time evolution.
The analytic setup is chosen such that the developed code is capable of following the evolution even inside the black hole region.
This, in particular, allowed us to investigate the mass inflation phenomenon in the chosen framework.
\end{abstract}

\pacs{04.20.-q, 04.25.-g}

\section{Introduction}

Relativistic infinitely thin spherical shells play an important role in various dynamical contexts ranging from microscopic to astrophysical systems.
For instance, by applying a charged shell as an electron model one may avoid the appearance of negative gravitational mass
caused by the concentration of charge at the center \cite{Lopez88,Zloshchastiev99a,Varela07}.
Using families of spherically symmetric thin shells instead of spherically symmetric continuous matter distributions
reduces significantly the complexity of evolutionary problems as the dynamics of thin shells may be investigated
by using various analogies from the description of the motion of a particle in a one-dimensional effective potential.
Accordingly, the quantization of systems comprising thin shells is tractable \cite{Dolgov97,Zloshchastiev99a,Corichi06,Ortiz07}.
Macroscopically stable quark-gluon matter can also be studied with a toy model
in which relativistic shells and the MIT bag model are combined \cite{Zloshchastiev99b}.
Collapsing dust shells can be used to probe stability or to study energetics of compact objects
such as back holes or star models mimicking the properties of black holes \cite{Ruffini03,Liu09,Gaspar10}.
Shells can be used to model matter ejection at certain phases of supernova explosions \cite{Ostriker71} or in modeling supernova remnants \cite{Zaninetti10}.
More realistic radiating shell models can also be constructed \cite{Hamity78} and with the help of these models,
even the critical collapse may be investigated analytically \cite{Koike95}.
With the help of infinitesimally thin shells one can construct exact solutions by gluing together spherically symmetric spacetime domains.
This way exotic models such as gravastars \cite{Visser04,Benedict05}
or wormholes \cite{Visser89,Visser95,Poisson95,Lobo04,Eiroa04,Dias10,Camera11} may also be studied.
Simple models of large-scale voids in galaxy distributions can also be constructed with the help of shells \cite{Maeda83,Lake85,Khakshournia01}.
The dynamics of spherical shells come into play in  some cosmological models, such as higher dimensional brane cosmologies,
in which it is assumed that our four-dimensional universe is merely a surface living in a higher dimensional spacetime \cite{Anchordoqui01}.
Shells play a central role in the bubble inflation model of the early universe \cite{Vilenkin84,Blau87,Berezin87,Farhi90,Sakai94}.
It is widely thought that by studying dynamics of shells, important phenomena such as the focusing singularity at the center \cite{Vick86}
or the so-called acoustic singularity---which will be discussed later in section 4.1.---can also be studied.

The basic equations governing the dynamics of thin shells can be derived in various ways,
for instance by making use of the junction conditions of the metrics \cite{Israel66,Nunez93,Goldwirth95,Nunez97,Nakao99,Goncalves02,Kirchner04,Krisch07,Krisch08},
by applying distributions \cite{Khorrami90,Mansouri96,Nozari02}
or the variational Jacobi--Hamiltonian approach \cite{Hajicek97,Ansoldi97,Gladush01,Crisostomo04,Kijowski05},
or by deriving them by taking a limit of the evolution equations of the thick shells \cite{Hoye85,Khakshournia02}.
It is important to mention that all of these methods produce the same set of basic equations.

In describing the evolution of families of infinitesimally thin shells, the study of their crossing is essential.
Nevertheless, much less has been done in this respect.
For instance, the basic equations describing shell crossing have only been derived for some specific cases,
such as totally transparent \cite{Fackerell75,Nakao99,Eid00} or totally inelastic collisions \cite{Langlois02,Gaspar10}.
Nevertheless, most of the authors do not go beyond deriving the equations of motion for dust shells,
or studying only the simplest possible analytic cases. This, in particular,
means that almost no results are available for multi-layer shell systems with the generic equation of state (EOS).
There are, apparently, only two papers in the literature which carry out a systematic study of the dynamics of more than two shells
or consider the evolution of shell systems such that shell crossings are allowed.
In \cite{Fackerell75} the dynamics of star clusters is studied,
although considerations therein remain on the theoretical side,
providing only the generic equations of motion in Schwarzschild time coordinates which, in particular,
does not allow the study of the motion through the event horizon.
In \cite{Eid00} the dynamics of two dust shells was investigated with the use of Kruskal--Szekeres coordinates---thereby
the authors were able to follow the motion of the system below the horizon---and shell crossings were assumed to be totally transparent.

Our main aim in this paper is to present some new results concerning the dynamics of multi-layer shell systems with the inclusion of of shell crossings.
The corresponding dynamical investigations were carried out by using a C++ code \cite{C++}
which made the study of the evolution of systems composed of a large number of shells and with a generic EOS possible.
In section 2, some of the basics related to the analytic description of the motion of a single shell are recalled
using Schwarzschild and ingoing Eddington--Finkelstein `time' coordinates.
For comparison, the corresponding Newtonian case is also discussed.
In section 3, for the sake of simplicity, first only the interaction of two shells is considered,
providing the balance equations for the case of totally transparent shell crossings.
In section 4, the time evolution of systems comprising a large number of concentric shells---in particular,
the formation of acoustic singularities---is investigated.
One of the main advantage of the method applied in this paper is related to the use of ingoing Eddington--Finkelstein coordinates,
which allows us to evolve even very complex composite systems throughout the entire spacetime, including the black hole region.
One of the most important outcomes of the investigations is that the phenomena of mass inflation
could also be studied by making use of colliding thin shells.
The paper is concluded by our final remarks.

Throughout this paper, the geometrized units are used, with $G = c = 1$,
and the abstract index notation of \cite{wald} will be applied with the additional use of
uppercase Latin indices signifying quantities living on three-dimensional hypersurfaces.

\section{The dynamics of a single shell}

In this section the basic equations relevant for a spherically symmetric infinitely thin shell,
bounded by two Schwarzschild vacuum regions, will be recalled.

\subsection{Equation of motion}
\label{emo}

Consider now a single shell and assume that the metric of the Schwarzschild regions on the sides of the shell is given as

\begin{equation}
\label{met1}
\rmd s_\pm^2 = - f_\pm(r)\,\rmd t_\pm^2 + f_\pm(r)^{-1}\rmd r^2 + r^2\,(\rmd\vartheta^2+\sin^2\hskip-.8mm{\vartheta}\,\rmd\varphi^2),
\end{equation}

\noindent
were the indices $\pm$ signify the outer and inner regions, respectively, $r$ stands for the area radius,
which is assumed to be continuous across the shell, $t_{-}$ and $t_{+}$ denote the Schwarzschild time coordinates in the corresponding spacetime regions,
$\vartheta$ and $\varphi$ are the standard spherical coordinates, while

\begin{equation}
\label{f}
f_\pm(r) = 1-\frac{2M_\pm}{r}
\end{equation}

\noindent
with mass parameters $M_\pm$, respectively\,\footnote{
Note that the form of the metric (\ref{met1}) with the slightly more generic metric function $f_\pm(r)=1-2 M_\pm(r)/r$ is suitable to represent
besides the Schwarzschild metric, that of the de Sitter spacetime in the vacuum case, or whenever electrovacuum spacetimes are also included
the Reisner--Nordstr\"om de Sitter solutions also fit this form.}.

Denote by $u^a$ the four-velocity tangent to the timelike generators of the shell
and by $\tau$ the proper time along these timelike generators.
The components of $u^a$ can be given as

\begin{equation}
\label{uup}
u^\alpha_\pm = \left(\frac{\rmd t_\pm}{\rmd\tau},\frac{\rmd r}{\rmd\tau},0,0\right)\,.
\end{equation}

\noindent
The induced metrics, $h_{AB}^-$ and $h_{AB}^+$, on the mutual boundary of the two spacetime regions are assumed to coincide,
and the metric on the shell $h_{AB} = h_{AB}^+ = h_{AB}^-$, in the $(\tau,\vartheta,\varphi)$ coordinates, reads

\begin{equation}
h_{AB}={\rm diag}(-1,{r}^2(\tau),{r}^2(\tau)\sin^2\vartheta)\,.
\label{hab}
\end{equation}

\noindent
As the four-velocity $u^a$ is of unit norm its components are not independent and,
in virtue of (\ref{met1}) and (\ref{uup}), the relation $u_{a}u^a = -1$ implies that the relation

\begin{equation}
\label{minus1}
\left(f_\pm \frac{\rmd t_\pm}{\rmd\tau}\right)^2 = f_\pm + \left(\frac{\rmd r}{\rmd\tau}\right)^2
\end{equation}

\noindent
holds from which, with the inclusion of the sign factor $\epsilon_{t_\pm}={\rm sign}(f_\pm \cdot {\rmd t_\pm}/{\rmd\tau})$,

\begin{equation}
\label{tdot}
\frac{\rmd t_\pm}{\rmd\tau} = \frac{\epsilon_{t_\pm}\sqrt{f_\pm(r)+(\rmd r/\rmd\tau)^2}}{f_\pm(r)}\,,
\end{equation}

\noindent
can be deduced. We shall return to the fixing of this sign factor later in subsection\,\ref{times}.
Nevertheless, let us mention here only that the value of $\epsilon_{t_-}$ or $\epsilon_{t_+}$ is $+1$
in regions where $t_-$ or $t_+$ is a timelike coordinate, i.e. above the respective horizons.

\begin{figure}
\center
\includegraphics[width=105mm,angle=0]{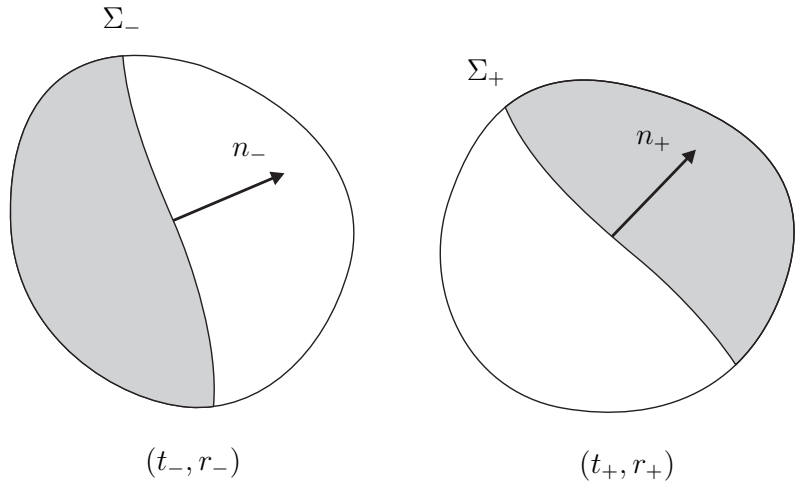}
\caption{The convention applied for fixing the sign of $\epsilon_{n_\pm}$ is illustrated.
The regions to be matched are indicated by the grey domains
while $\Sigma_{\pm}$ denotes their `isometric' boundaries in the preliminary spacetimes.}
\label{figure1}
\end{figure}

The unit normal $n_a$ to the shell, satisfying the orthogonality requirement $u^a n_a=0$, can be given as

\begin{equation}
n_\alpha{}_\pm=\epsilon_{n_\pm}\left(-\frac{\rmd r}{\rmd\tau},\frac{\rmd t_\pm}{\rmd\tau},0,0\right)\,,
\label{npm}
\end{equation}

\noindent
where the value of the sign factor $\epsilon_{n_\pm}$ is chosen such that $n_a$ points outward as it is illustrated in figure \ref{figure1}.
In the particular Schwarzschild case this implies that $\epsilon_{n_\pm}=+1$.\,\footnote{Note that by allowing $\epsilon_{n_\pm}$
to take the value $\pm1$ within the very same setup, wormholes \cite{Visser95,Poisson95} and other type of exotic spacetimes can be studied.
However, in this paper attention will be restrected to conventional shells.}

Now, with the help of the induced metric, $h_{AB}$, and the unite normal $n_a{}_\pm$ the extrinsic curvature tensors,
at the boundaries $\Sigma_-$ and $\Sigma_+$, are given as

\begin{equation}
K_{AB}^\pm=\frac12 {h_A}^E{h_B}^F \pounds_{n_\pm}h_{EF}\,.
\label{Kab}
\end{equation}

As already mentioned the equation of motion for the shell separating the Schwarzschild regions can be derived in various ways.
One of the most frequently applied methods is based on Israel's thin-shell formalism \cite{Israel66}
in which junction conditions are specified at the location of the shell.
In particular, besides the relation $h_{AB}=h_{AB}^+=h_{AB}^-$, the discontinuity of the pertinent extrinsic curvature tensors,
$K_{AB}^+$ and $K_{AB}^-$, is assumed to be related to the surface energy-momentum tensor of the shell
$S_{AB} = \mathrm{diag}(\sigma,{\mathcal P}\hskip-1mm\cdot\hskip-1mm r^2,{\mathcal P}\hskip-1mm\cdot\hskip-1mmr^2\sin^2\theta)$---where
$\sigma$ and $\mathcal{P}$ denote the surface energy density and the two-dimensional tangential pressure of the shell,
respectively---via the Lanczos equation

\begin{equation}
\label{matchcond}
K_{AB}^+-K_{AB}^-=-8\pi\left\{S_{AB}-\frac12 h_{AB}\left(h^{CD}S_{CD}\right)\right\}\,.
\end{equation}

\noindent
By making use of (\ref{npm}) and (\ref{Kab}) we get $K_{\vartheta\vartheta}^\pm=\epsilon_{n_\pm} r f_\pm ({\rmd t_\pm}/{\rmd\tau})$.
Then, in virtue of (\ref{matchcond}), the $\vartheta$-$\vartheta$ component of this junction condition can be seen to take the form

\begin{equation}
\label{master}
s_-\sqrt{f_-(r)+\left(\frac{\rmd r}{\rmd \tau}\right)^2}
- s_+\sqrt{f_+(r)+\left(\frac{\rmd r}{\rmd \tau}\right)^2} = 4\pi\sigma r\,,
\end{equation}

\noindent
where $r$ now signifies the radius at the location of the shell,
while the sign factors $s_-$ and $s_+$ are nothing but the products of $\epsilon_{t_\pm}$ and $\epsilon_{n_\pm}$,
i.e. $s_\pm=\epsilon_{t_\pm} \epsilon_{n_\pm}$.
Since $\epsilon_{n_\pm}=+1$, $s_\pm=\epsilon_{t_\pm}$ hereafter.

Note that the importance of the appropriate treatment of the sign factors $s_-$ and $s_+$ was emphasized in \cite{Goldwirth95,Fayos95}.
Regardless of which of the alternative methods (mentioned in the introduction) is applied,
the reasonings always ends up with the equation of motion

\begin{equation}
\label{rdot}
\left(\frac{\rmd r}{\rmd \tau}\right)^2 = \left(\frac{\MG}{\MR}\right)^2 - 1 + \frac{2\MC + \MG}{r} + \left(\frac{\MR}{2r}\right)^2\,,
\end{equation}

\noindent
where $\MC=M_-$ denotes the Schwarzschild mass parameter of the central region,
$\MR = 4\pi\sigma r^2$ is the `rest mass' of the shell representing the surface internal energy associated with the tangential motion of particles,
while $\MG=M_+-M_-$ is the `gravitational mass' of the shell.
Accordingly, $\MC$ and $\MG$ are constants but in general $\MR$ is a function of the radius,
and it is constant only in the particular case of dust shells with zero pressure.

To determine the $r$-dependence of $\MR$ we need an additional equation which can be derived by making use of the conservation law $D^AS_{AB}=0$,
where $D_A$ denotes the covariant derivative associated with the metric $h_{AB}$.
In this particular case the conservation law takes the form \cite{Nakao99}\,\footnote{Equation
(\ref{conserv}) may also be derived by making use of the two algebraically independent components of (\ref{matchcond}).}

\begin{equation}
\label{conserv}
\frac{\rmd}{\rmd\tau}\left(\sigma r^2\right) =
\mathcal{P}\frac{\rmd}{\rmd\tau}\left(r^2\right)\,,
\end{equation}

\noindent
By introducing the area radius $r$, instead of $\tau$, as our independent variable (\ref{conserv}) can be put into the form

\begin{equation}
\label{dsigma}
r\frac{\rmd\sigma}{\rmd r} = -2(\sigma + \mathcal{P})\,.
\end{equation}

\noindent
This is the point where the EOS of the shell comes into play.
With the help of an EOS of the form $\mathcal{P} = \mathcal{P}(\sigma)$ the functional form of $\sigma=\sigma(r)$ may be determined,
which, in turn, gives us the functional form of $\MR(r)$, as well.
Since (\ref{dsigma}) is a separable differential equation an implicit solution to it---provided that
the EOS is regular enough to guarantee the above integral to exist---can be written as

\begin{equation}
\label{EOSsolution}
\frac{r}{r_0} = \exp{\left(-\frac{1}{2}\int_{\sigma_0}^{\sigma(r)}
\frac{\rmd \tilde{\sigma}}{\tilde{\sigma} + \mathcal{P}(\tilde{\sigma})}\right)}\,,
\end{equation}

\noindent
with integration constant $\sigma_0=\sigma(r_0)$.

Once $\MR(r)$ is known from (\ref{rdot}), along with suitably chosen initial conditions,
can be used to determine the motion as a function of the proper time $\tau$.

\subsection{Equation of state}

Although the developed C++ code allows the use of basically any kind of EOS in the numerical investigations covered by this paper,
only the homogeneous linear EOS is applied. It is also important to keep in mind
that in specifying an EOS some additional restrictions always need to be taken into account.
For instance, the use of a suitable energy condition is essential.
The most appropriate one is the so-called dominant energy condition (DEC) guaranteeing that solutions to dynamical problems respect the concept of causality.
For a selected type of infinitesimally thin shell DEC can be seen to hold whenever $|\mathcal{P}| < \sigma$,
which, in particular, means that $\sigma$ is non-negative (see, e.g., \cite{Gaspar10,Barcelo02}).
To avoid dynamical instabilities we shall also assume that the square of the speed of sound,
$\CS^2=\rmd \mathcal{P}/\rmd\sigma$, is non-negative, and, to be compatible with the conventional concept of relativity,
that $\CS^2$ is less than or equal to the square of the speed of light.

As mentioned above, for the shake of simplicity, in this paper considerations will be restricted to the simplest possible functional form,
i.e. to a homogeneous linear EOS $\mathcal{P}(\sigma)=w\sigma$ in which case (\ref{EOSsolution}) takes the form

\begin{equation}
\label{EOS1}
\sigma(r) = \sigma_0\,\left(\frac{r}{r_0}\right)^{-2(1+w)} \,.
\end{equation}

\noindent
DEC, along with hydrodynamical stability on the surface,
can be seen to hold whenever $w$ is chosen from the interval $0\leq w\leq 1$.
Note also that $w=0$ corresponds to the dust case.

\subsection{Initial conditions}
\label{InCon}

In applying Israel's junction conditions one may start by fixing the geometrical properties of the spacetimes on the respective sides of the shell,
and then solving the junction conditions for suitable rest mass and initial velocity values making the matching possible.
This could be referred to as a `mathematician approach',
while in a `physicist approach' one would start by fixing the rest mass and initial velocity of the shell
and then try to determine suitable geometrical parameters of the respective side spacetime regions.
The latter approach is applied below.

Accordingly, in describing the motion of a shell we regard $\MC$ as the environmental parameter
and $m_0,\, r_0,\, v_0$ as initial parameters of the shell at $\tau_0$.
Here $m_0=\MR(r_0),\, r_0=r(\tau_0),\, v_0=\dot{r}(\tau_0)$,
and the over-dot denotes derivative with respect to the proper time $\tau$.
As mentioned above, to be compatible with DEC we shall assume that $m_0>0$.
In virtue of equation (\ref{rdot}), or (\ref{master}), whenever the junction is possible the gravitational mass, $\MG$,
depends only on the kinetic energy and it may be expressed in terms of the initial data as

\begin{equation}
\label{mg}
\MG = m_0\left(-\frac{m_0}{2r_0} + \sqrt{1 - \frac{2\MC}{r_0} + v_0^2}\,\right)\,.
\end{equation}

In order to avoid the gravitational mass to becoming complex the inequality

\begin{equation}
\label{notcomplex}
v_0^2 \geq \frac{2\MC}{r_0} - 1\,
\end{equation}

\noindent
has to hold.

In virtue of (\ref{mg}) $\MG$ may be negative.
Note also that since $\sigma$ is required to be positive $\epsilon_{t_-}=-1$ and $\epsilon_{t_+}=+1$ cannot occur.
As we shall see below, (\ref{et}) excludes the possibility of having both $\epsilon_{t_-}$ and $\epsilon_{t_+}$ be negative.
Thus, in the remaining two cases, the sufficient conditions ensuring the existence of the desired type of matching are given as

\begin{table}[h]
\begin{center}
\begin{tabular}{|c|c|c|l|}
\hline
$\epsilon_{t_-}$ & $\epsilon_{t_+}$ & sufficient conditions &  gravitational mass \\
\hline
+ & + & $v_0^2 > A$ &  $\MG > 0$ \\
\hline
+ & $-$ & $v_0^2 < A$ &  $\MG > 0$, if $v_0^2 > B$; \;\, $\MG \leq 0$,  elsewhere \, \\
\hline
\end{tabular}
\label{table}
\end{center}
\end{table}

\noindent
where
\begin{eqnarray}
\label{A}
A = \frac{m_0^2}{r_0^2} + \frac{2\MC}{r_0} - 1\,, \\
\label{B}
B = \frac{m_0^2}{4 r_0^2} + \frac{2\MC}{r_0} - 1\,.
\end{eqnarray}

\subsection{Characterizing the dynamics}
\label{chardyn}

For dust shells the values of $\MG$ and $\MR$---the latter is also constant for dust shells---characterize the motion in the following way.
The system is said to be gravitationally bound whenever there exists a finite radius, $\RMAX$, such that the velocity vanishes at $\RMAX$.
In general, the value of $\RMAX$ coincides with the positive root of the right-hand side of (\ref{rdot}) which,
whenever $\MG < \MR$ reads as\,\footnote{It follows from (\ref{rmax}) that $\RMAX \geq R_{\mathrm{S}}=2(\MC+\MG)$ as mentioned before.}

\begin{equation}
\label{rmax}
\RMAX = \left(1-\frac{\MG^2}{\MR^2}\right)^{-1} \left(\MC + \frac{\MG}{2} + \sqrt{\MC^2 + \MC\MG + \frac{\MR^2}{4}}\,\right).
\end{equation}

If $\MG = \MR$ the kinetic energy becomes zero exactly at the spatial infinity,
i.e. whenever $\RMAX = \infty$, and the associated motion is usually referred to as `marginally  bound'.
For $\MG > \MR$ the motion of the shell is not restricted, and, in virtue of (\ref{mg}), at the spatial infinity,
i.e. in the $r_0\rightarrow \infty$ limit, the relation $\MG = \MR\sqrt{1+v_{\infty}^2}$ holds.

Although the solution to (\ref{rdot}) is, in  general, complicated, for dust shells it can be given in closed form (see, e.g, \cite{Hoye85}).
For example, in the special case of marginally bound motion, with $\MG=\MR$, the pertinent solution can be given by the implicit relation

\begin{equation}
\label{taur}
\tau(r) = \left.\frac{(4\tilde{r}\MC + 2\tilde{r}\MR - \MR^2)\sqrt{8\tilde{r}\MC + 4\tilde{r}\MR + \MR^2}}{24\MC^2  \MR^2}\right|_{\tilde{r}=r}^{\tilde{r}=\RI}\,,
\end{equation}

\noindent
where $\RI$ denotes the initial location of the shell at $\tau=0$.
Setting $\RI=0$ the value of $-\tau(r)$ becomes equal to the proper time duration meanwhile the shell collapses from radius $r$ to the symmetry center.
Specializing even further, for a `self-gravitating' shell with a Minkowski interior,
i.e. with $\MC=0$, the proper time that is needed for such a shell of radius $r$ to undergo a complete gravitational collapse can be given as

\begin{equation}
\label{tauc}
\tau_\mathrm{c}(r) = \frac{\MR}{6} + \frac{\sqrt{4r + \MR}}{3}\left(\frac{r}{\sqrt{\MR}}-\frac{\sqrt{\MR}}{2}\right).
\end{equation}

\noindent
It is worth keeping in mind that the above analytic expressions were derived by making use of the assumption that the motion is marginally bound which,
in particular, means that the velocity of the shell at $r$ should be as if the shell started to move towards the center from rest at spatial infinity.

Figure 2 shows some numeric examples for dust shells.
For shells with non-zero pressure, the situation is more complex because pressure may, and in fact does,
significantly alter the motion of the particles.
By investigating the case of a homogeneous linear EOS it was justified,
in contrast to the dust case, that such a shell may be in equilibrium,
although the pertinent equilibrium was found to be unstable \cite{Kijowski06,Goncalves02}.
A more complicated EOS such as polytrop can only be studied numerically, and---as was justified by our numerical experiences---it is possible
to construct shells which oscillate in a bounded region or which are in stable equilibrium.

\subsection{The Schwarzschild time and the Eddington--Finkelstein null coordinates}
\label{times}

Up to this point all the derivatives in the equations relevant to the evolution of the investigated shell were expressed
with respect to the proper time associated with the shell.
In many practical cases it turns out to be necessary---as it will clearly be demonstrated in the following sections
concerning the collision of shells---to know what is the functional relation of this proper time,
e.g., to the Schwarzschild time coordinates defined on the inner and outer sides of the shell.
We would like to emphasize that in the applied formalism only the area radius, $r$,
is required to be a continuous---although, not necessarily monotonic---function through the shells.
Accordingly, in general the Schwarzschild time coordinates defined on the sides need not match continuously.
Therefore, we have to determine the functional relation of the proper time to both the inner and outer Schwarzschild time coordinates separately.

In doing so recall that the derivative $({\rmd t_\pm}/{\rmd\tau})$ has been given by (\ref{tdot}).
Note, however, that the value of $\epsilon_{t_\pm}$---for its definition see subsection\,\ref{emo}---is undetermined yet,
although it is uniquely determined in regions where the Schwarzschild time coordinates increase in the future direction,
i.e. whenever the radius is larger than both of the Schwarzschild-radii,
where $\rmd t_\pm / \rmd \tau > 0$ and $f_\pm > 0$, the use of $\epsilon_{t_\pm}=+1$ in (\ref{tdot}) is required.

Nevertheless, as it was shown in \cite{Nakao99} the derivative $({\rmd t_\pm}/{\rmd\tau})$ can always be determined uniquely.
More concretely, by making use of the $\tau-\tau$ component of (\ref{matchcond}) it can be shown---for details see,
e.g., the part of section 2 in \cite{Nakao99} between equations (2.21) and (2.34)---that
regardless of the location of the shell, whenever it moves in a Schwarzschild spacetime
\begin{eqnarray}
{\label{t1}
\frac{\rmd t_{-}}{\rmd \tau}  =
\left(1 - \frac{2\MC}{r}\right)^{-1}\left(\frac{\MG}{\MR} + \frac{\MR}{2r}\right)\,,} \\
\label{t2}
{\frac{\rmd t_{+}}{\rmd \tau} =
\left(1 - \frac{2(\MC + \MG)}{r}\right)^{-1}\left(\frac{\MG}{\MR} - \frac{\MR}{2r}\right)\,,}
\end{eqnarray}

\noindent
or equivalently, the relations ${\epsilon_{t_\pm}\sqrt{f_\pm(r)+(\rmd r/\rmd\tau)^2}} = {\MG}/{\MR} \mp {\MR}/{2r}$ are always satisfied everywhere.
This, in particular, implies that

\begin{equation}
\label{et}
\epsilon_{t_\pm}=\mathrm{sign}({\MG}/{\MR} \mp {\MR}/{2r})\, .
\end{equation}

It follows from (\ref{et}) that either $\epsilon_{t_-}$ or $\epsilon_{t_+}$ changes sign depending on the sign of $\MG$.
This sign change occurs at the `critical radius' $\widehat r=\MR^2/(2 |\MG|)$.\,\footnote{Note that whenever ${\mathcal P}\geq 0$
the value of $\widehat r$ is unique due to the monotonicity of the functions $\MR=\MR(r)$.
It can also be checked that unless this change of the sign occurs, the worldsheets representing the evolving shells
cannot be of class $C^2$ at $r=\widehat r$.}

Once the initial values for $t_{-}$ and $t_{+}$ are specified, these equations determine the desired relations
between the proper time and the Schwarzschild time coordinates in the inner and outer spacetime regions, respectively.
The differences of the right-hand sides of (\ref{t1}) and (\ref{t2}) make it clear that,
apart from very exceptional cases, even though the initial values for $t_{-}$ and $t_{+}$
are chosen to coincide, they will necessarily differ latter, i.e. they need not match continuously as was indicated above.

The Schwarzschild metric has a coordinate singularity at the event horizon,
i.e. at $r=2\MC$ in the inner region and at $r=2(\MC+\MG)$ in the outer region.
The Schwarzschild time coordinate becomes infinite while approaching the horizon (see for example figure 2);
thereby, it does not allow the proper description of the motion through the horizon.
It is possible to overcome this technical difficulty by applying ingoing Eddington--Finkelstein coordinates
covering both the domain of outer communication and the black hole regions simultaneously.
The ingoing Eddington--Finkelstein null coordinate is defined as $\rm{v}_\pm={t_\pm}+r^{*}$,
where $r^{*}$ denotes the `tortoise coordinate' determined by the relation $\rmd r^{*}/\rmd r = f_\pm^{-1}(r)$.
Accordingly,

\begin{equation}
\label{null0}
\frac{\rmd \rm{v}_\pm}{\rmd\tau} = \frac{\rmd t_\pm}{\rmd\tau} + \frac{\rmd r^{*}}{\rmd r}\frac{\rmd r}{\rmd\tau} =
\frac{\epsilon_{t_\pm}\sqrt{f_\pm(r)+(\rmd r/\rmd\tau)^2} + \rmd r/\rmd\tau}{f_\pm(r)}\,,
\end{equation}

\noindent
where $\epsilon_{t_\pm}$ is given by (\ref{et}).

\begin{figure}
\center
\includegraphics[width=54mm,angle=270]{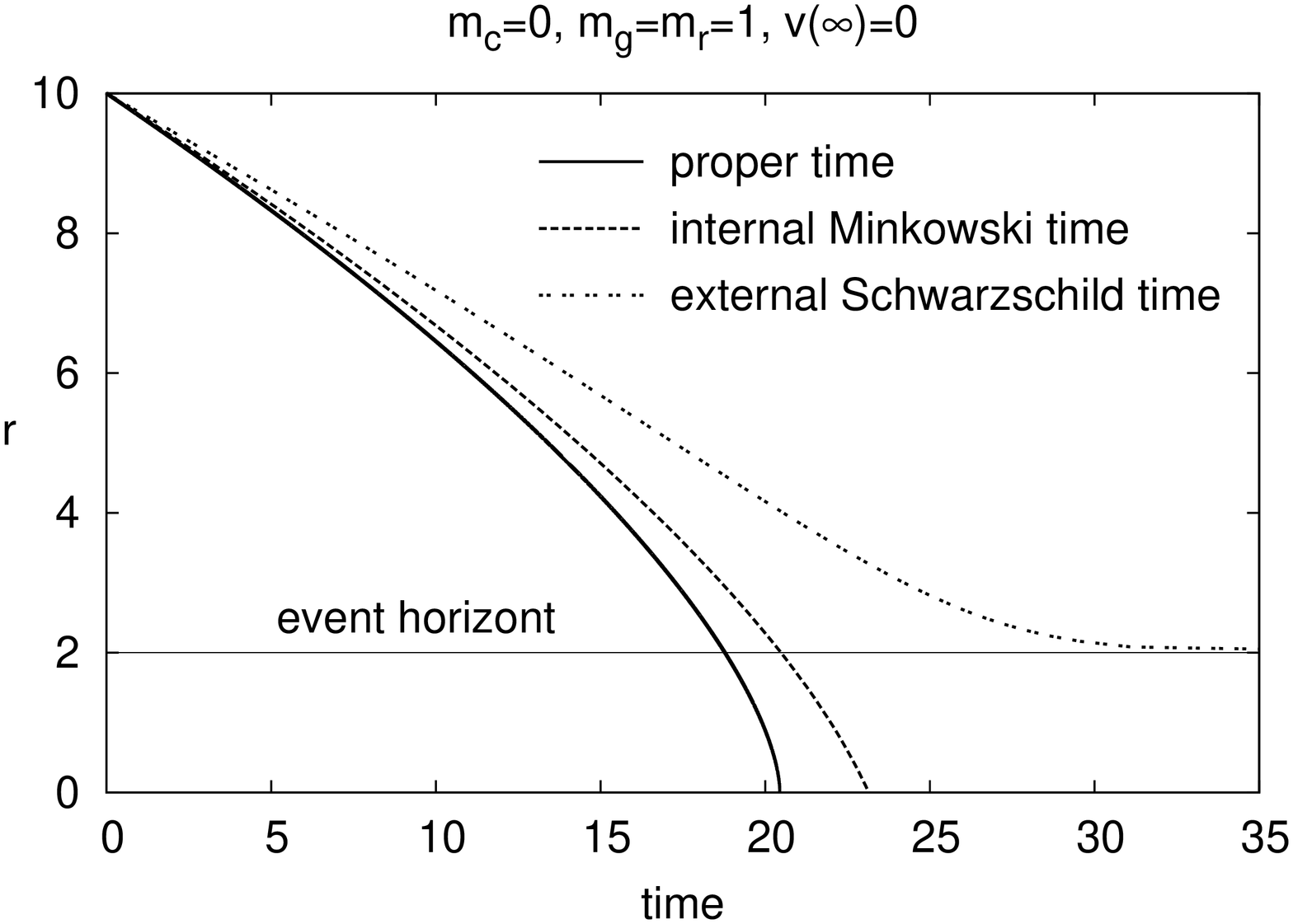}
\includegraphics[width=54mm,angle=270]{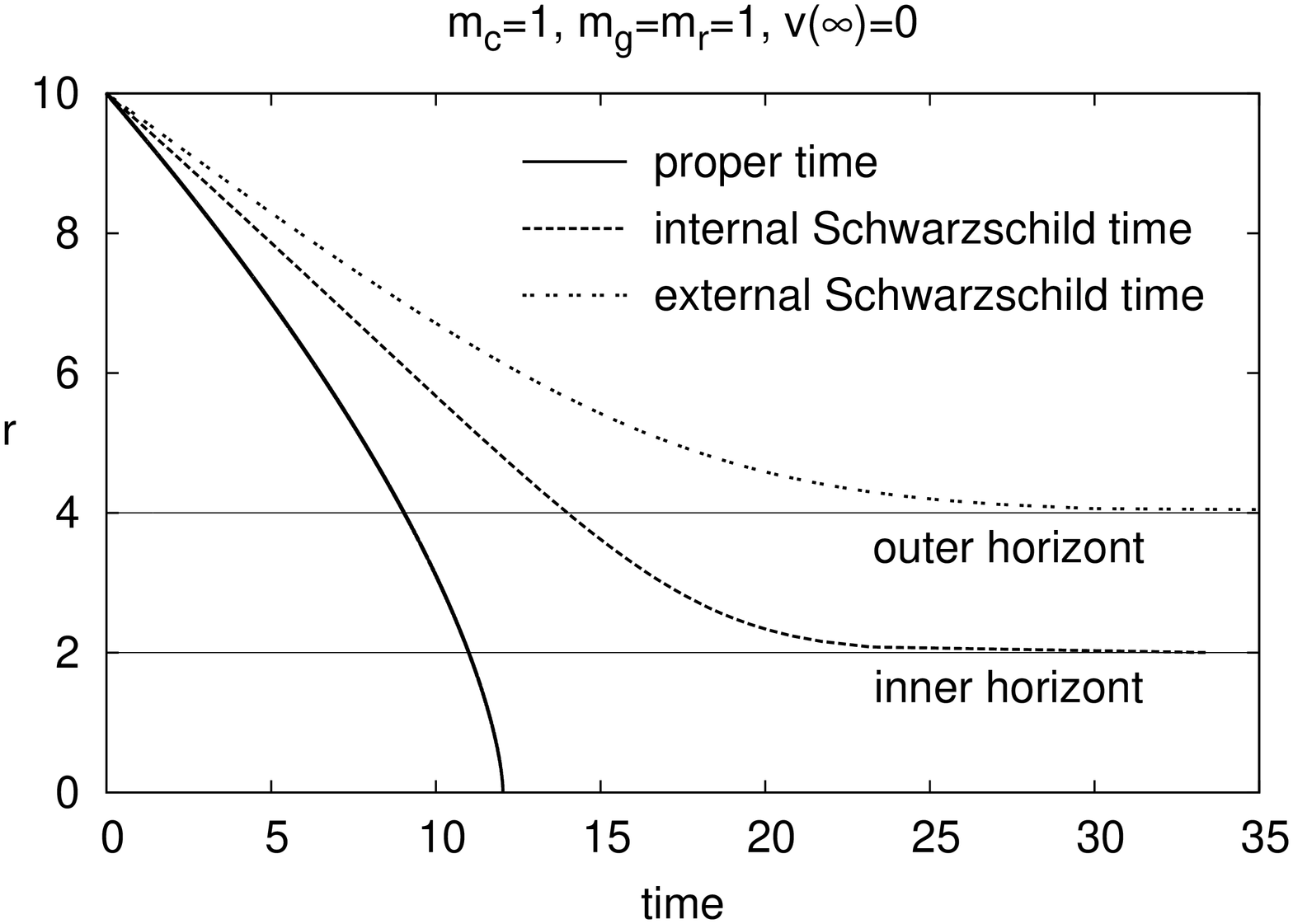}
\caption{
In the left panel the motion of a single dust shell, starting at $r=10$, in the particular case with $\MC=0,\ \MR=\MG=1$ is shown
with respect to the proper time, the interior Minkowski time and the exterior Schwarzschild time.
In the last case, as is indicated by the pertinent plot, the shell can only reach the event horizon asymptotically,
while the time of complete collapse is $\tau_\mathrm{c}=(19\sqrt{41}+1)/6\approx 20.44$
or $\TC=(11\sqrt{41}-1)/3\approx 23.14$ with respect to the proper or inner Minkowski time, respectively.
In the right panel the motion of a similar shell with $\MC=\MR=\MG=1$ is indicated.
In this case the inner and outer horizons denote the horizons at the central and exterior Schwarzschild regions, respectively.
}
\label{figure2}
\end{figure}

Note that in the particular case of a dust shell with $\MC=0$ and $\MG=\MR$, by making use of (\ref{t1}) and (\ref{tauc}),
the time needed for a complete gravitational collapse may be given in terms of the interior Minkowski time as

\begin{equation}
\label{minktc}
\TC(r) = \frac{\sqrt{4r + \MR}}{3}\left(\sqrt{\MR} + \frac{r}{\sqrt{\MR}}\right) - \frac{\MR}{3}\,.
\end{equation}

In concluding this section we would like to emphasize that in describing the motion of a single shell---determined by (\ref{rdot})---only
the mass parameters of the interior and exterior Schwarzschild spacetime regions,
along with the mass of the shell (determined by the surface energy density and the pertinent EOS of the shell), are relevant.
In particular, as long as there is no collision between a selected and the surrounding shells,
the motion of the selected one depends only on the mass parameters of the interior and exterior Schwarzschild regions
Note also that, likewise in the Newtonian case, the mass of the interior shells (if they exist at all) comes into play only via the value of the central mass $\MC$.
However, as the gravitational mass $\MG$ may not be positive, the general relativistic motion of the shell
may differ significantly from that of a shell, possessing the same $\MC$ and $\MR$ parameters, in the Newtonian theory.

\subsection{The motion of shells in the Newtonian case}

In this section, for the sake of completeness and for comparison,
a brief review of the Newtonian framework will be given first.
Then, the equations of motion will be derived for fluid shells.
Throughout this subsection the basic quantities in the Newtonian framework will be signified by uppercase letters
corresponding to the ones introduced in the previous sections for fully relativistic systems.

Note first that in the Newtonian theory the inertial mass and the gravitational mass of a shell are not distinguished.
This common mass of the shell---which, in the Newtonian description, is independent of time and the EOS---will be denoted by $M_{\rm shell}$.
The other considerable simplification characterizing the Newtonian theory comes from the use of absolute time, denoted by T.

Likewise in the fully relativistic case the equation of motion of the shells in the Newtonian framework can be derived in various ways.
In the case of dust shells, i.e. for shells with zero pressure, a good review can be found in \cite{Frauendiener95}.
According to the argument outlined therein the basic equation is nothing but a balance equation of force per unit mass
and for the dust case it possesses the form of (\ref{newtonian1}) below with $P=0$.

In generalizing this result to the fluid shells,
i.e. in determining the functional form of the last term on the right-hand side of (\ref{newtonian1}) with $P\not=0$,
we need to find the appropriate force term representing the contribution of the pressure to the Newtonian balance equation.
In identifying it let us consider an elementary `square shaped' surface element of the shell with sides $R\,\Delta \phi $,
where $\Delta \phi$ denotes the viewing angle of the sides from the center.
The mass of this elementary piece is $\Delta M_{\rm shell} = \Sigma\,R^2 (\Delta \phi)^2$,
where $\Sigma=M_{\rm shell}/(4\pi R^2)$ is the surface mass density.
There are four elementary forces exerted at the sides of this square-shaped surface element.
The size of these elementary forces is $\Delta F=P\,R\,\Delta \phi$,
where $P$ denotes the `two-dimensional' pressure of the shell.
Since the directions of these forces are tangential to the shell, the resultant total force is radially outward pointing
and its size is $\Delta F_{\mathrm{tot}} = 4\Delta F\sin{(\Delta \phi/2)}$.
Correspondingly,  $\Delta F_{\mathrm{tot}}$ tends to $2\Delta F\Delta \phi$ in the $\Delta \phi \rightarrow 0$ limit which, in turn,
justifies that the force per unit mass, due to the non-zero pressure,
is $\lim_{\Delta \phi \rightarrow 0}\Delta F_{\mathrm{tot}}/\Delta M = 8\pi P R/M_{\rm shell}$.
Accordingly, the equation of motion is of the form

\begin{equation}
\label{newtonian1}
\ddot{R} = -\frac{2M_{\mathrm{c}}+M_{\rm shell}}{2R^2} + \frac{8\pi P R}{M_{\rm shell}}
\end{equation}

\noindent
and the initial condition to this differential equation consists of $M_{\rm shell}$, $R_0$ and $V_0$,
where $V_0=\rmd R/\rmd T$ at $R_0$. In addition, as in the relativistic case, we need the central mass $M_{\mathrm c}$ as an environmental variable.
Formally the gravitational mass can be calculated from the initial conditions just as in the Einstein theory,
but, as we noted above, in this case the gravitational mass has to coincide simply with $M_{\rm shell}$.
In comparing results relevant for the Newtonian and Einstein theories we need to harmonize initial conditions.
Note that while $r_0$ and $R_0$, or $v_0$ and $V_0$, have identical interpretations in both theories,
in setting the value of $M_{\rm shell}$ we have the following inequivalent two choices.
We may identify $M_{\rm shell}$ with either the rest mass or the gravitational mass in the Einsteinian setup.
As in the Newtonian regime, i.e. whenever $r_0\gg \MC,\MG$ and $v_0\ll 1$, the relation $\MG\sim\MR$ holds,
we could identify either of them with $M_{\rm shell}$. Nevertheless, as the gravitational mass plays the same role conceptionally
in both theories, it seems to be more appropriate to assume the equality of the two gravitational masses by setting $M_{\rm shell}=\MG$.

Assuming that $P$ is a given function of the radius $P=P(R)$, the equation of motion, (\ref{newtonian1}), can be integrated.
The existence of such a function is always guaranteed whenever an EOS of the form $P=P(\Sigma)$ is known because $\Sigma=M_{\rm shell}/(4\pi R^2)$
itself is a function of the radius. It is worth emphasizing that the EOS was kept as completely generic throughout the above discussion.

Multiplying (\ref{newtonian1}) by $\dot{R}$ and by then integrating with respect to $T$ one gets,
as the  analog of (\ref{rdot}), the Newtonian energy balance equation

\begin{equation}
\label{newtonian2}
\dot{R}^2 = V_0^2 + \left[\frac{2M_{\mathrm{c}}+M_{\rm shell}}{\tilde{R}}\right]_{R_0}^R + 4\int_{R_0}^{R}\frac{W(\tilde{R})}{\tilde{R}}\rmd \tilde{R}\,,
\end{equation}

\noindent
where $W(R)=P(R)/\Sigma(R)$. In particular,
in the case of a shell with a linear EOS where $W(R) = \overline W=const$,
the last term on the right-hand side of (\ref{newtonian2}) takes the form $4{\overline W}\ln{(R/R_0)}$.

\section{The evolution of multi-layer shell systems}
\label{evoshells}
\setcounter{equation}{0}

This section provides a brief review of the analytic and numerical setup we have applied in studying the evolution of multi-layer shell systems.
This evolution, in general, involves a large number of collisions of the shells.

\subsection{Collision of two shells}
\label{collshells}

As it was emphasized in section IV of \cite{Frauendiener95},
the description of a collision of two shells cannot be given without invoking some further assumptions
concerning the interaction of the shells. More concretely,
there is an ambiguity in the evolution of colliding shells even though the energy and momentum conservations are guaranteed to hold as---depending on
the type of interactions of the shells---more than two shells or even a continuous spread of the matter of the original two shells
into a thick shell might develop during the collision.

\begin{figure}[ht]
\vspace{1mm}
\center \unitlength 0.85mm
\begin{picture}(60,60)(-30,-30)
\thicklines
\put(0,0){\line(-1,-1){25}}
\put(0,0){\line(1,-1){25}}
\put(0,0){\line(1,1){25}}
\put(0,0){\line(-1,1){25}}
\put(-25,-27){\makebox(0,0)[t]{Shell 1}}
\put(25,-27){\makebox(0,0)[t]{Shell 2}}
\put(25,27){\makebox(0,0)[b]{Shell 3}}
\put(-25,27){\makebox(0,0)[b]{Shell 4}}
\put(-20,0){\makebox(0,0){Region 1}}
\put(0,-18){\makebox(0,0){Region 2}}
\put(20,0){\makebox(0,0){Region 3}}
\put(0,18){\makebox(0,0){Region 4}}
\end{picture}\hspace{2cm}
\begin{picture}(60,60)(-30,-30)
\thicklines
\put(0,0){\line(-1,-1){25}}
\put(0,0){\line(1,-1){25}}
\put(0,0){\line(0,1){25}}
\put(-25,-27){\makebox(0,0)[t]{Shell 1}}
\put(25,-27){\makebox(0,0)[t]{Shell 2}}
\put(0,27){\makebox(0,0)[b]{Shell 3}}
\put(-20,11){\makebox(0,0){Region 1}}
\put(0,-18){\makebox(0,0){Region 2}}
\put(20,11){\makebox(0,0){Region 3}}
\end{picture}
\vspace{1mm}
\caption{
Schematic spacetime diagrams representing the totally transparent (left) and totally inelastic (right) collisions.
The vertical direction is temporal and time progresses upward.}
\label{figure3}
\end{figure}
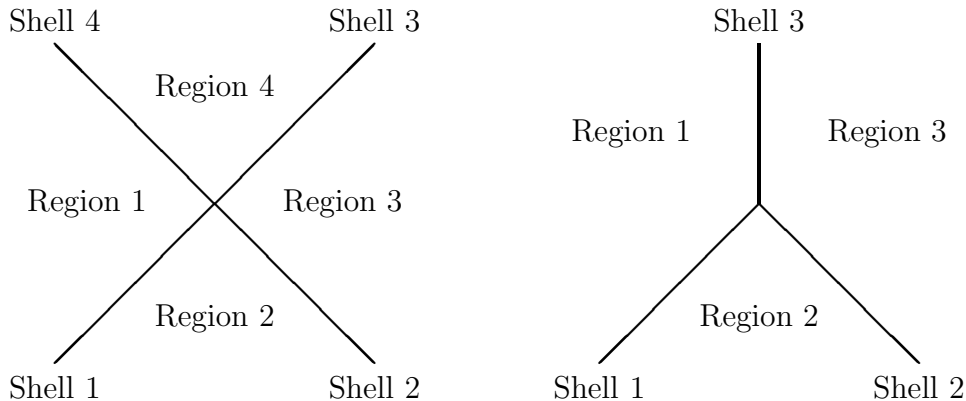

This ambiguity is eliminated if the two shells pass through each other either without any interaction,
in which case the collision is said to be {\it totally transparent},
or when the interaction is extremely strong and the ingoing shells merge into a single outgoing shell,
in which case the collision is referred to as {\it totally inelastic}.
These two extreme cases are schematically represented in figure \ref{figure3}.
In this paper only the totally transparent shell crossings will be considered.
The EOS of the shells will also be assumed to be intact in collisions.
Note that while for dust shells this assumption seems to be appropriate
it is much less adequate whenever the shells are comprised of strongly interacting particles.
Below, the basic equations relevant for totally transparent collisions are recalled.\,\footnote{A detailed analytic
description of the totally inelastic case has been given in the appendix of \cite{Gaspar10}
while the C++ code \cite{C++} is developed so that it is capable of investigating
the evolution of shell systems when the collisions are assumed to be totally inelastic.}

Detailed investigation of the dynamics of the totally transparent collision of two shells can be found in \cite{Nakao99}.
For this type of collision it is assumed that $\MR{}_3=\MR{}_1$ and $\MR{}_4=\MR{}_2$ and---provided that
the four-velocity of each shell is continuous at the location of collision---the momentum conservation can be shown
to be equivalent to the relations \cite{Nakao99}
\begin{eqnarray}
\label{transparentp3}
p_3 & = & p_1 + \Delta p \,, \\
\label{transparentp4}
p_4 & = & p_2 + \Delta p \,,
\end{eqnarray}

\noindent
where $p_i = \MR{}_{i} (\rmd r/\rmd \tau)_{i}$ stands for the 3-momenta of the $i${th} shell---here
the indexing of the shells follows the notation applied on the left-hand side of figure \ref{figure3}---and

\begin{equation}
\Delta p = -\frac{\MR{}_{1}\MR{}_{2}}{\RC{}} u^a_1n_2{}_a=\frac{(\MG{}_{1}-h_1)\,p_2-(\MG{}_{2}+h_2)\,p_1}{\RC{}-2(\MC{}_{1}+\MG{}_{1})}\,,
\end{equation}

\noindent
where $\RC{}$ is the radius at the collision, $\MR{}_{i}$ and $\MG{}_{i}$ stand for the rest mass and the gravitational mass of the $i${th} shell,
$\MC{}_{1}$ denotes the central mass, while $u^a_i$, $n^a_i$ and $h_i=\MR{}_{i}^2/(2\RC{})$ denote the four-velocity,
the unit normal and the `self-gravity' of the $i${th} shell, respectively.

Note that in (\ref{transparentp3}) and (\ref{transparentp4}) both of the positive signs in front of the term $\Delta p$ are correct as,
in order to suit to the three-momentum conservation, $\Delta p$ itself is negative.
The apparent conflict may be resolved immediately if one takes into account that $p_3$ and $p_4$,
as well as $p_1$ and $p_2$, are components of four-momentum vectors with respect to different bases \cite{Nakao99}.
It also follows from $\Delta p<0$ that, in order to have a transparent collision the vector fields $u^a_1$ and $n^a_2$ have to be arranged so
that the contraction $u^a_1n_2{}_a$ be positive.

From the four-momentum conservation the `energy balance' relations
\begin{eqnarray}
\label{transparentp5}
&& \MG{}_3=\MG{}_1-\Delta \mathcal{E}\,, \\
\label{transparentp6}
&&\MG{}_4=\MG{}_2+\Delta \mathcal{E} \,,
\end{eqnarray}

\noindent
can also be derived, where $\MG{}_1=M_2-M_1$, $\MG{}_2=M_3-M_2$, $\MG{}_3=M_3-M_4$, $\MG{}_4=M_4-M_1$;
the $M_{i}$ stands for the mass parameter of the $i${th} Schwarzschild region as indicated on the left panel of figure \ref{figure3}
and the measure of the `energy transfer', $\Delta \mathcal{E}$, can be given as

\begin{equation}
\Delta \mathcal{E}=-\frac{\MR{}_{1}\MR{}_{2}}{\RC{}} u^a_1u_2{}_a\,.
\end{equation}

\noindent
Since both $u^a_1$ and $u^a_2$ are future-directed timelike vectors the contraction $u^a_1u_2{}_a$ is negative
implying that $\Delta \mathcal{E}$ is always positive. This, in particular, means that
the outer shell loses energy, while the energy of the inner shell increases.

In subsection \ref{minf} the above relations will be economized in characterizing the mass inflation phenomenon.

Note that in the case of multi-layer shell systems one should take into account the possibility of simultaneous collisions of more than two shells.
Since the occurrence of these type of events is practically zero,
by choosing the time steps to be sufficiently small we could guarantee that in each time step merely pairwise collisions occurred.
It is also important to note that whenever a simultaneous collision of more then two shells is allowed to occur
the result may always be determined by decomposing the event into pairwise collisions
and the result is independent of the order of the pairing process.

Note, finally, that in the Newtonian case the basic equations for totally transparent collisions are simple as
the masses and the velocities are interchanged in the most obvious way.

\subsection{Dynamics of multi-layer shell systems}

In describing the relative motion of multi-layer shell systems we need to specify a reference parameter.
As long as considerations are restricted to two shells,
the most convenient reference parameter is the Schwarzschild time coordinate defined in the intermediate region
while we remain in the region of outer communication,
whereas inside the black hole region the Eddington--Finkelstein null coordinate serves the same purpose.
Let us mention that without choosing an appropriate reference parameter
one may create significant confusion even in the (simplest possible) case of two shells.
For instance, in \cite{Goncalves02} where the respective proper times measured along the separate shells---these,
in virtue of (\ref{t1}) and (\ref{t2}), may differ considerably---were assured to coincide and were used as a reference parameter,
false conclusions were derived concerning, e.g., the extent of regions where the crossing of shells may occur.

In describing the motion of multi-layer systems consisting of $N\,(>2)$ shells,
labeled by the index $i\in \{1,...,N\}$, the Schwarzschild time coordinates defined in the respective intermediate regions,
associated with any pair of shells next to each other, are---as long as only the two shells are concerned---as suitable as before\,\footnote{If
the motion is intended to be described inside the event horizon in these intermediate regions
Eddington--Finkelstein null coordinates have to be applied instead of the Schwarzschild time
as it has been indicated several times. Note also that, because of spherical symmetry,
we frequently replace the spacetime by its factor space with respect to the group SO(3).}.
However, as was discussed before, these Schwarzschild time coordinates cannot be matched properly due to their discontinuities across the shells.
To overcome this technical difficulty the following synchronization process may be applied.
In each of the intermediate regions the constant $t$-lines determine simultaneity.
The entire spacetime--built up from piecewise Schwarzschild regions---is synchronized by applying this synchronization process
to succeeding intermediate regions starting from the innermost region.

By making use of this synchronization method, the location, i.e. the $r^{(i)}$ coordinates of all the shells,
can be given as a function of the time coordinate of the innermost region,
playing the role of reference parameter, $t_r$.
In most of the following figures instead of plotting the $r^{(i)}(t_r)$ functions,
the expressions $\Delta r^{(i)}(t_r)=r^{(i)}(t_r)-\bar r(t_r)$,
where $\bar r(t_r)$ denotes the radial center of mass of the $r^{(i)}(t_r)$ distribution,
i.e. $\bar r(t_r)=\sum_{i=1}^N \MR^{(i)}r^{(i)}(t_r)/\sum_{i=1}^N \MR^{(i)}$,
will be plotted\,\footnote{By simply reversing the orientation of the synchronization,
i.e. by starting from the outside we may also use the Schwarzschild time coordinate of the external region as our reference time.}.
Interestingly, in certain cases, e.g.\,in case of a collapsing shell system
where $\bar r=\bar r(t_r)$ is guaranteed to be a monotonically decreasing function,
instead of $t_r$ the radial center of mass $\bar r$ can also be used as a reference parameter.
For instance, in some of the figures (see, e.g., figures \ref{figure4}, \ref{figure5}, \ref{figure7} and \ref{figure8})
the functions $\Delta r^{(i)}=\Delta r^{(i)}(\bar r)$ will be plotted.

It is important to keep in mind that the above-defined synchronization has a certain degree of ambiguity
concerning the specific $t_r$-value at a given point.
Nevertheless, once the synchronization is fixed, the crossings of shells are well-defined as the $t_r$-labels are uniquely determined,
i.e.\,the motion of the shells may be properly represented in the $(t_r,r)$ local coordinates.
Note also that using the Schwarzschild time coordinates in the intermediate regions we may follow the motion of the shells
only up to the appearance of the event horizon in either of these regions.
One could argue that an external observer cannot see what happens to the shells beyond an the event horizon,
which manifest itself by the infinite value of the Schwarzschild time coordinate.
Nevertheless, in some cases, such as in the study of mass inflation discussed in subsection \ref{minf},
it is important to descend into the black hole region. Fortunately,
just like in the case of a single shell, the use of the Eddington--Finkelstein null coordinates
instead of the Schwarzschild time coordinates resolves the corresponding problem of synchronization for multi-layer systems.

\subsection{Initial data for systems}

In specifying the initial data for a shell system, just like in the simple shell case,
we need as an environmental parameter the central (Schwarzschild) mass, $m_\mathrm{S}$, characterizing the innermost region.
The rest of the initial data set consists of the triples $m_{0}^{(i)}$, $r_{0}^{(i)}$, $v_{0}^{(i)}$ for the individual shells,
with $i\in\{1,2,...,N\}$, which are required to satisfy the relations $r_{0}^{(i)} < r_{0}^{(j)}$ for $i<j$.
In determining the gravitational mass, $\MG^{(i)}$, of the $i$th shell---by making use of the pertinent form of (\ref{mg})---we need to know
the central mass, $\MC^{(i)}$, felt by the $i$th shell. It can be justified that
for $i>2$ the relation $\MC^{(i)}=\MC^{(i-1)}+\MG^{(i-1)}$ holds, with $\MC^{(1)}=m_\mathrm{S}$.
Note that the initial data $m_{0}^{(i)}$, $r_{0}^{(i)}$, $v_{0}^{(i)}$ also have to satisfy the pertinent form of (\ref{notcomplex})for each individual shell.

Note also that, in specifying the initial state of our multi-layer shell system the EOS of the individual shells have to be fixed.

\subsection{The numeric algorithm}

The open source C++ code of the applied numerical algorithm can be downloaded (see \cite{C++}).
This subsection is to provide a short outline of this numerical algorithm.

Apart from shell collisions, the motion of the individual shells can be treated separately when we suppress the indices of shells hereafter.
In order to be able to integrate the equations of motion---see (\ref{rdot})---we need to know $\MC$, $\MG$ and $\MR(r)$.
Once $\MC$ and suitable initial data---consisting of $m_0$, $r_0$ and $v_0$---are specified
the value of $\MG$ and $\MR(r)$ can determine as on one hand $\MG$ is given by (\ref{mg}) while on the other hand,
$\MR(r)=4\pi\sigma(r) r^2$ and the functional relation $\sigma=\sigma(r)$ can be deduced with the help of an EOS,
as was described at the end of subsection \ref{emo}. For instance, in the particular case of a dust or fluid shell
with a homogeneous linear EOS, $\MR(r)$ can be given analytically by making use of (\ref{EOS1}).
In all the other more generic cases numerical algorithms can always be used in solving (\ref{EOSsolution}).

Once $\MR(r)$ is known, the absolute value of $\rmd r/\rmd\tau$, along with $(\rmd t_\pm/\rmd \tau)$, can be determined
as a function of the radius by making use of (\ref{rdot}), (\ref{t1}) and (\ref{t2}).
The sign in front of the velocity $\rmd r/\rmd\tau$ depends on the initial data and its value remains fixed
until the appearance of a turning point of the shell if it occurs at all.
The velocity, with respect to the Schwarzschild times $\rmd r/\rmd t_\pm$,
can also be given by applying the chain rule

\begin{equation}
\label{vchain}
\frac{\rmd r}{\rmd t_\pm} = \frac{\rmd r}{\rmd \tau} \left(\frac{\rmd t_\pm}{\rmd
\tau}\right)^{-1}\,.
\end{equation}

\noindent
In describing the motion of a multilayer shell system, proceeding from the inside to the outside,
we need to apply the synchronization outlined above. Therefore, the motion of each shell is determined
with respect to both the inner and outer Schwarzschild time (or the Eddington--Finkelstein null coordinates)
because---as follows from the synchronization process---in specifying the `time step' in the outer regions we need to know
the `time step' applied in the innermost region.

In integrating the equations of motion the fourth-order Runge--Kutta algorithm was applied.
The appearance of a turning point in the integration process is indicated by the fact that the right-hand side of (\ref{rdot}) becames negative.
When this happens the code does not undertake the last step and by making use of a root finding
the location of the turning point is determined. The equation of motion (\ref{rdot}) is then integrated
such that the sign of the velocity, ${\rmd r}/{\rmd \tau}$, calculated from it is changed to the opposite of its previous value.
A change in the order of the shells, i.e. the ordering of their radial coordinates,
always indicates that a collision occurred somewhere. Then the evolution of the system is then held on,
for a short while, until the precise location (in space and time) of the collision is determined by linear approximation.
At that event, by making use of the conservations equations and the assumption about the EOS,
we determined the initial data for the new shell(s).
(If inelastic collisions are allowed to occur the number of shells has to be decreased by 1.)
The evolution of the full multi-layer system is carried on with the inclusion of the yielded new shell(s) afterwards.
Once the innermost shell reaches the origin, the evolution of the system is continued so that it is taken out of the system of shells
and its gravitational mass is added to the center mass of the innermost region.
Whenever the Eddington--Finkelstein coordinates were applied, the algorithm was basically the same
with the distinction that derivatives $\rmd t_\pm/\rmd \tau$ were replaced with the corresponding expressions $\rmd \rm{v}_\pm/\rmd \tau$.

The precision of the applied numerical schema is checked in the case of a system
formed by two repeatedly intersecting equal rest mass dust shells (see figure \ref{figure11} in section \ref{minf}).
This justifies that our numerical code is convergent even though succeeding collisions occur.

\section{The main results}
\setcounter{equation}{0}

One of the most interesting questions in working with systems of thin shells is how to mimick thick shells.
However tempting certain analogies might be, one should keep in mind that even though one is using a huge number of thin shells,
the continuum distributions cannot be properly modeled, with the exception of the dust case,
because in the thin shell limit the interaction of particles in the direction transversal to the shells---apart from the gravitational one---is neglected.
What is possible to do consistently is to investigate an approximate model using a large number of dust shells.
Let us mention here that the only attempt to investigate such an approximate model---as far as we know---was made in \cite{Eid00}
although the number of the shells was kept minimal, i.e.\,only the evolution of a two-shell system was investigated.
In contrast to this, with the help of our C++ code \cite{C++},
the evolution of a large number of shells can be determined.

\subsection{The study of `simple' systems}

In order to provide a better understanding of the dynamics of our thick shell mimicking
the multi-layer thin shell system as reference solutions,
the evolution of $N=16$ shells with uniform initial mass and radial distributions will first be examined.
Since even in this case there are too many parameters characterizing the system,
it seems to be appropriate---as will be done below---to proceed in small steps by changing the initial conditions
of the reference solution almost parameter by parameter.

\begin{figure}
\center
\vspace{-4.5mm}
\includegraphics[width=51mm,angle=270]{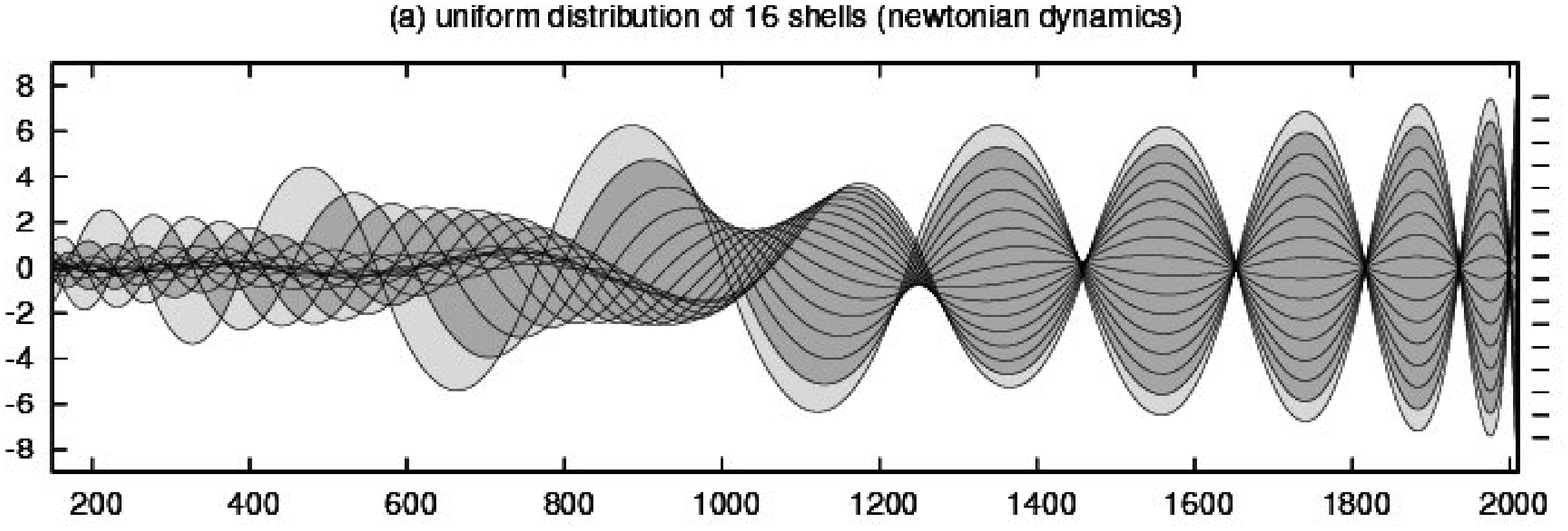}\\
\includegraphics[width=51mm,angle=270]{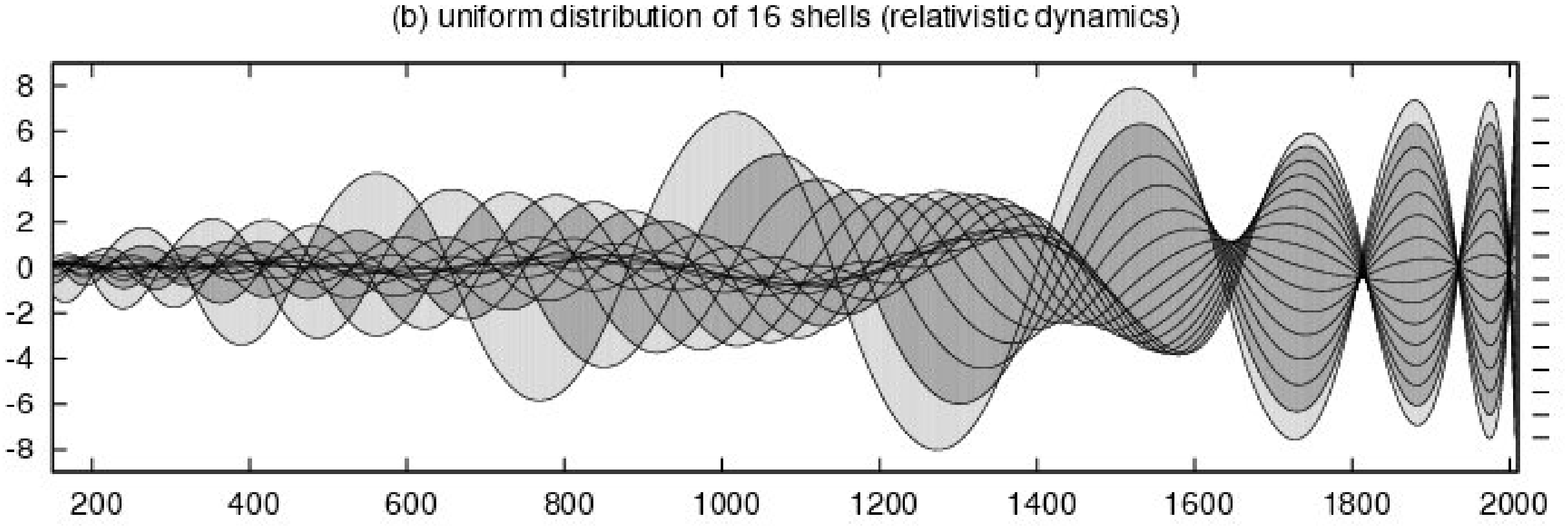}\\
\includegraphics[width=51mm,angle=270]{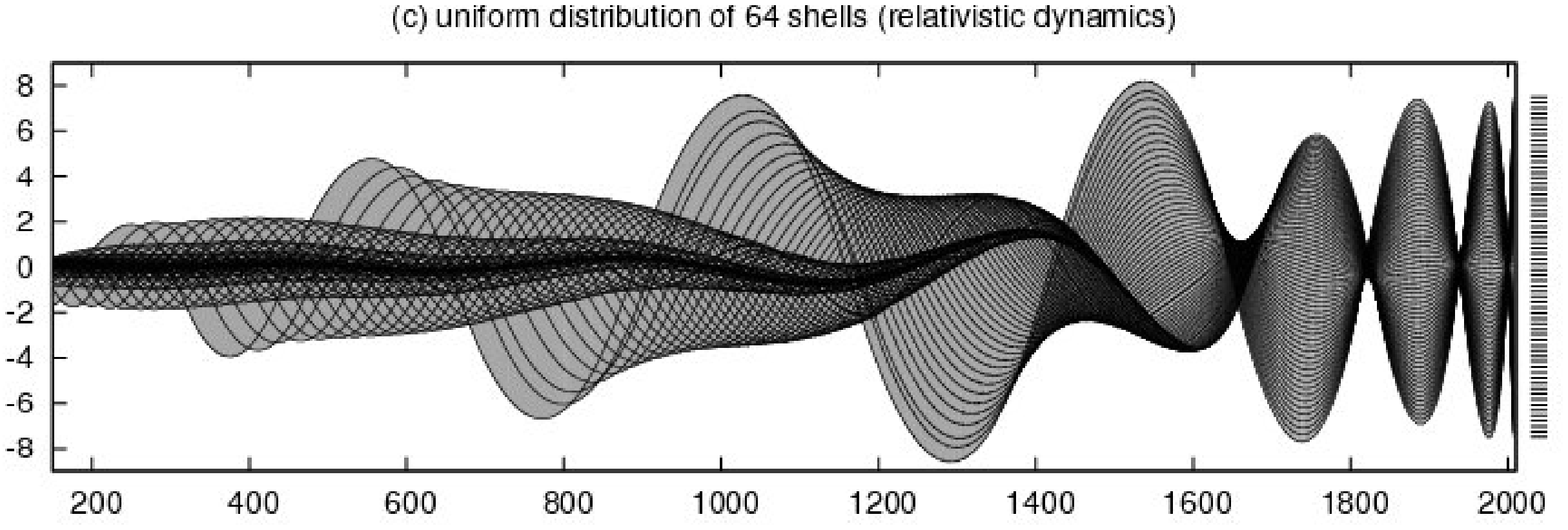}\\
\includegraphics[width=51mm,angle=270]{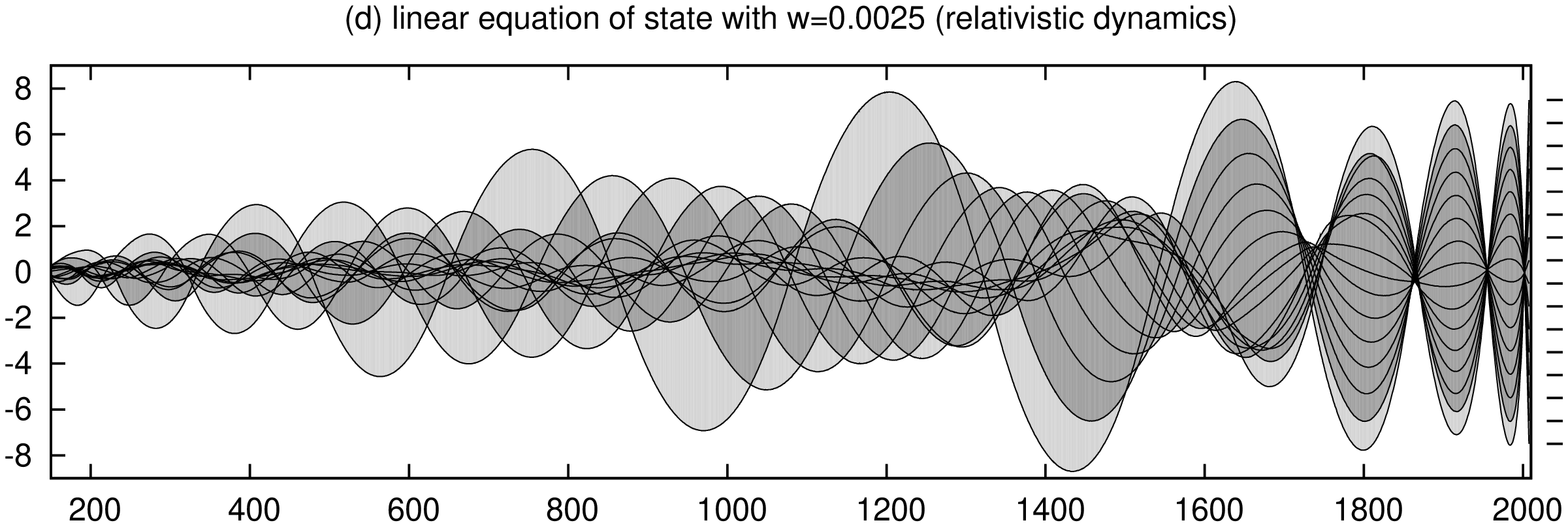}\\
\caption{The time evolution of shell systems with uniform mass and initial radial distributions is shown.
On the horizontal axis the radial center of mass, $\bar r$, of the radius of the shell system is indicated,
while on the vertical axis the deviation $\Delta r^{(i)}=r^{(i)}-\bar r$ relevant for the individual shells is plotted.
The upper two panels are to indicate the similarities and differences between the Newtonian and fully relativistic evolutions.
Although it might look strange at first, it is important to keep in mind that for the depicted collapsing systems---where $\bar r$
is monotonically decreasing---here, and in figures \ref{figure5}, \ref{figure7}, \ref{figure8}, \ref{figure9} and \ref{figure10},
time progresses from right to left.}
\label{figure4}
\end{figure}

\begin{figure}
\center
\vspace{-3.5mm}
\includegraphics[width=51mm,angle=270]{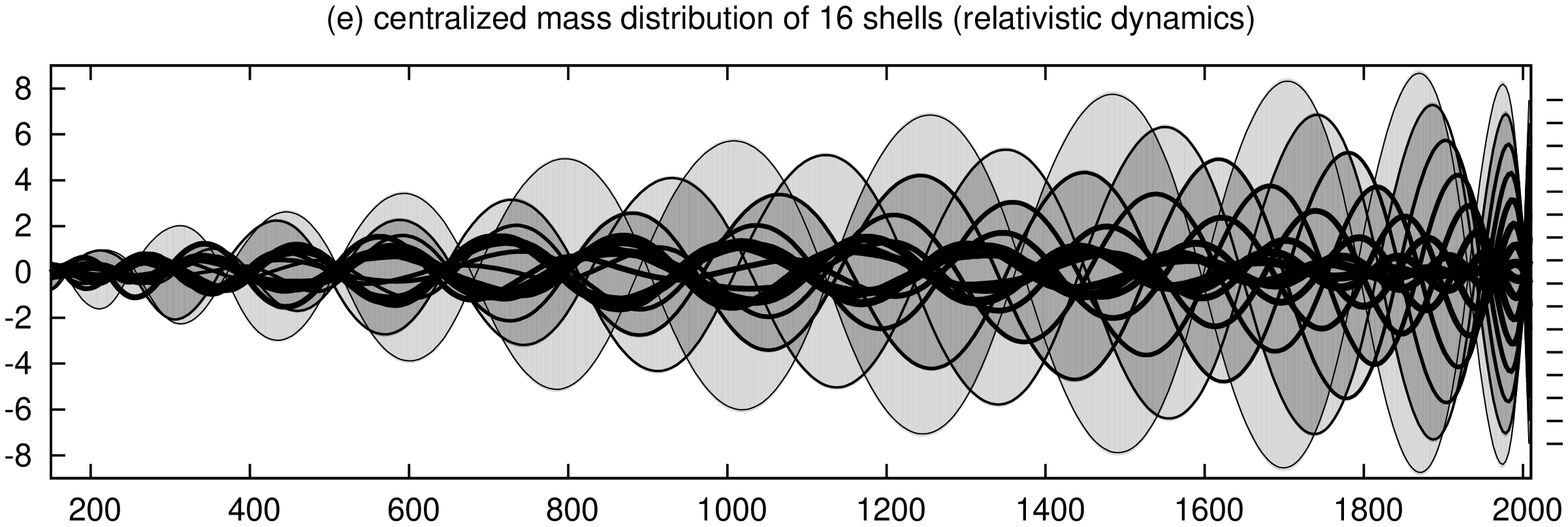}\\
\includegraphics[width=51mm,angle=270]{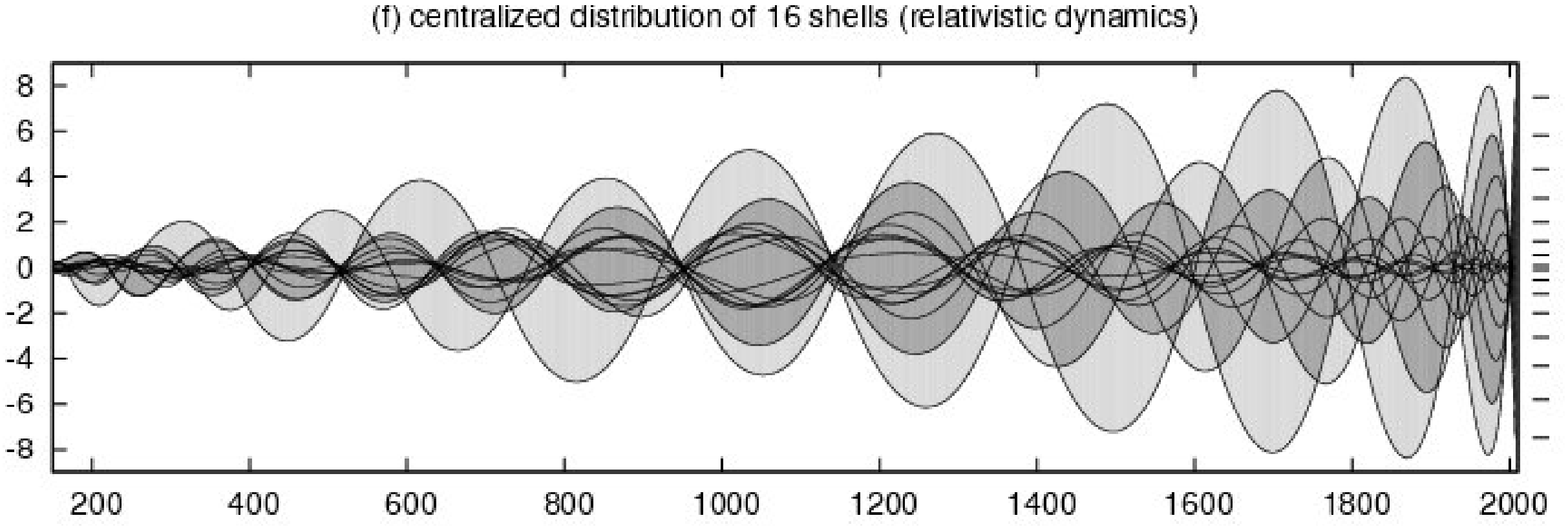}\\
\includegraphics[width=51mm,angle=270]{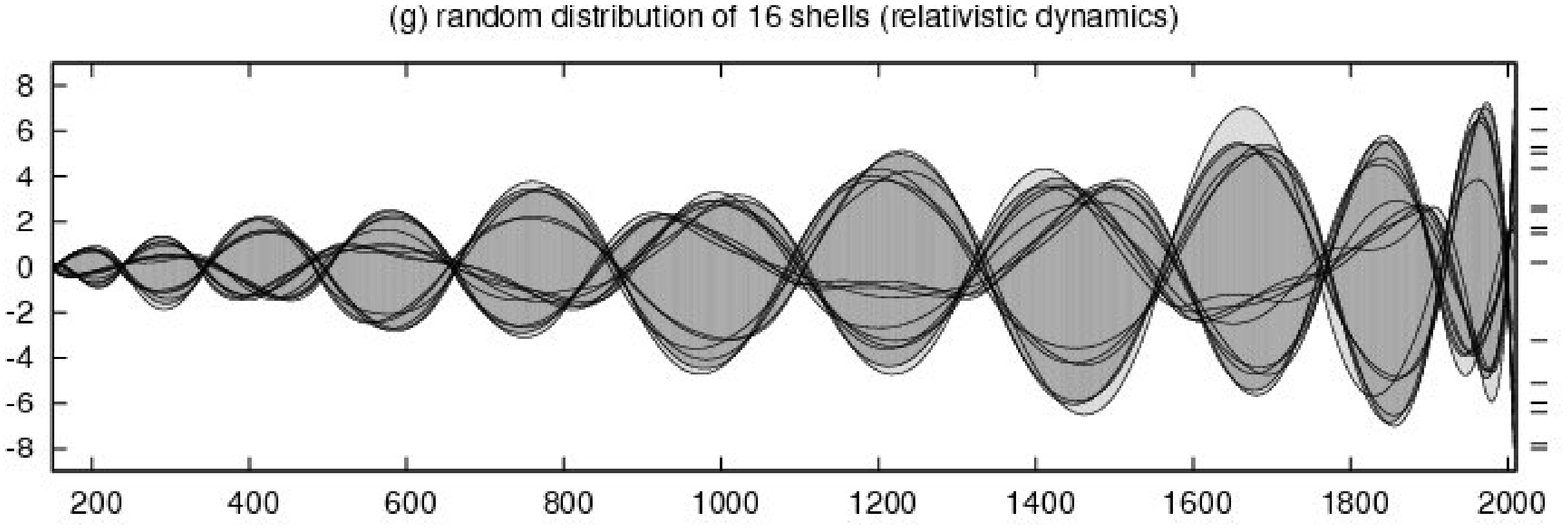}\\
\includegraphics[width=51mm,angle=270]{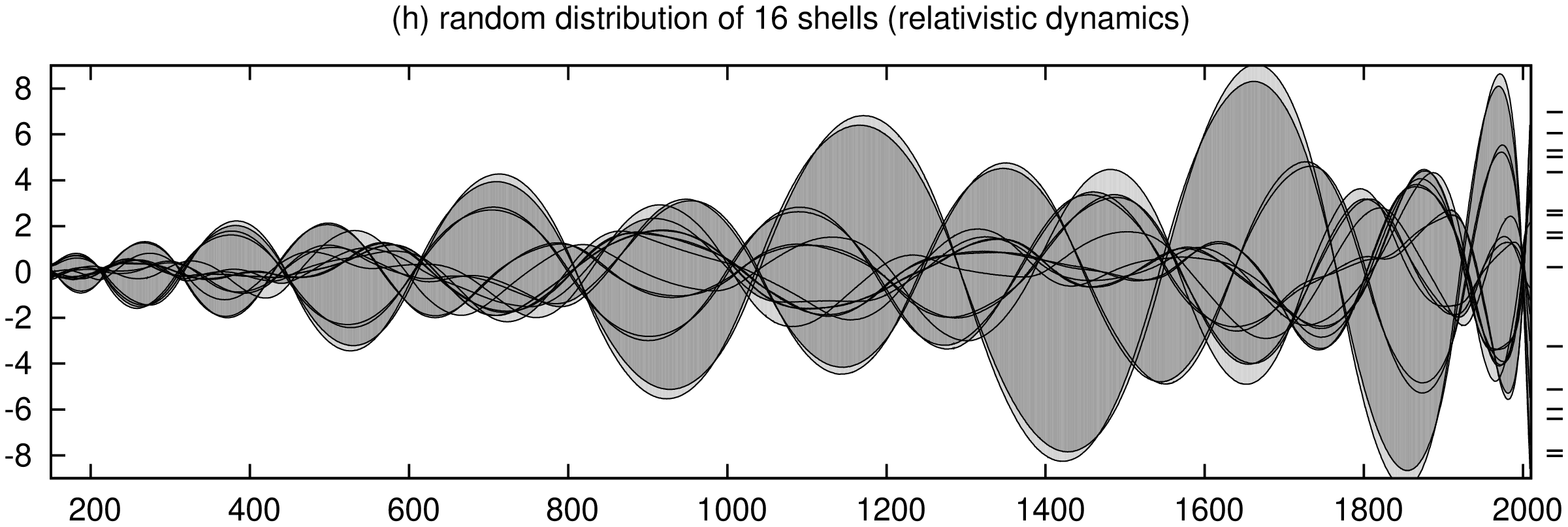}\\
\caption{The time evolution of shell systems are shown.
In the top panel the mass distribution is uniform while in the other three panels the initial radial distribution is non-uniform;
in fact, the initial radial distribution is chosen to be random on the lower two panels.
As in figure \ref{figure4} the radial center of mass, $\bar r$, of the radius of the shell system is indicated on the horizontal axis,
while on the vertical axis the deviation $\Delta r^{(i)}=r^{(i)}-\bar r$ relevant for the individual shells is plotted.
In spite of the significant differences in the initial part of the indicated evolutions,
the similarities in the final parts are considerable.}
\label{figure5}
\end{figure}

As mentioned above we shall consider the time evolution of shell systems consisting of $N=16$ shells,
with the exception of panel (c) in figure \ref{figure4} where the evolution of $N=64$ shells will be considered.
The total rest mass $\sum_{i=1}^{N} m_0^{(i)}$  of these shells will be $72$, while $m_\mathrm{S}=0$,
in each of the following cases. The mass distribution will be uniform on each panel of figure \ref{figure4},
while it is uniform on panel (e) and centered on panel (f), possessing the mass distribution as specified by (\ref{mmm}).
The initial radial distribution $r_0^{(i)}$, in the case of uniform distributions,
is chosen so that $r_0^{(i)} = 1999+i$, where $i$ runs from 1 to 16.
It is centered on panel (f) and it is random on panels (g) and (h) in figure \ref{figure5}.
It is important to keep in mind that even in the case of a uniform mass distribution,
only the rest masses $m_0^{(i)}$ could be arranged to be equal to each other,
whereas the corresponding surface mass densities $\sigma^{(i)}_0$ differ slightly according to the relations $m_0^{(i)}=4\pi \sigma{}^{(i)}_0 r_0{}^{(i)}$.
For simplicity, only dust shells are considered---with the exception of panel (d) in figure \ref{figure4}---and the initial velocity,
$v_0^{(i)}$, was chosen to be zero, i.e. all the shells start from rest.

On each of the panels of figures \ref{figure4} and \ref{figure5} on the horizontal axis the radial center of mass $\bar r$ is indicated,
which means that the synchronized events are arranged so that they lie along vertical segments.
The initial radius distribution of the shells is indicated by strokes on the right side of each plot.
In order to help the recognition of the developing structures the following type of gray shadowing had been applied.
A lighter gray shade was used between the momentary outermost and the last but one shell,
while the complementary inner part received a darker gray coloring.
Since we intended to compare the Newtonian and fully relativistic time evolutions in the relativistic case,
all the plots were produced by making use of the Schwarzschild time coordinates,
whence all the evolution stops before reaching `the even horizon'.

Let us provide now a brief outline of the plots in figures \ref{figure4} and \ref{figure5}.
The letters applied in the following itemization refer to the labels of the the individual panels.

\begin{itemize}

\item[(a)] In the Newtonian setup the time evolution of the above-specified shell system with an uniform radius and mass distribution is shown.

\item[(b)] The time evolution, with initial data corresponding to that of panel (a),
is depicted in the fully relativistic case.

\item[(c)] This justifies that the basic characters do not change drastically whenever the numbers of the shells is increased by a factor of 4.
The time evolution of 64 shells is depicted in the fully relativistic case.

\item[(d)] The fully relativistic time evolution of 16 shells is considered using the same initial configuration as in panel (b)
with the distinction that the EOS of the shells is homogeneous linear, ${\mathcal P}=w\sigma$ with $w=0.006$.\,\footnote{The order of magnitude
of the value of $w$ was determined as if the shell system formed a virialized system in the kinetic gas theory.}

\item[(e)] In this plot the evolution of a system with a uniform initial radial distribution but with the centralized

\begin{equation}
\MR^{(i)} = \left\{
\begin{array} {c l}
i , & {\rm if}\  1 \leq i \leq 8;\\
16-i , & {\rm if}\  9 \leq i \leq 16\,.
\end{array}
\right.\label{mmm}
\end{equation}

\noindent
mass distribution is shown. The latter is indicated by the thickness of the horizontal strokes at the right edge of each of the individual figures.

\item[(f)] In this panel the evolution of a system, with uniform initial mass distribution and with symmetrically centered radial distribution,
is shown, where $r_0^{(1)}=2000$, $r_0^{(16)}=2015$ and
\begin{eqnarray}
&&r_0^{(i+1)}-r_0^{(i)}=(9-i)\cdot\Delta\,,\ \ \rm{for}\ \  i=1,2,...,7 \,; \\
&&r_0^{(i)}-r_0^{(i-1)}=(i-8)\cdot\Delta\,,\ \  \rm{for}\ \  i=10,11,...,16\,,
\label{rrr}
\end{eqnarray}

\noindent
with $\Delta=15/71$.

\item[(g),(h)] In these two panels the mass distribution is uniform while the initial radial distributions are uniform random distributions,
although the initial location of the outermost shells were chosen to be the same as on all the other plots, i.e.\,$r_0^{(1)}=2000$ and $r_0^{(16)}=2015$.

\end{itemize}

By comparing the plots in figures \ref{figure4} and \ref{figure5} the following conclusions may be drawn.
It is clearly visible that whenever the shells are starting from rest, not only the nearby shells cross but the crossing of all the shells seems to be generic.
It has been justified by our numerical experiences--see also panel (c) in figure \ref{figure4}---that
the developing basic structures seem to be independent of the number of shells.
Let us mention that shell crossings are known to occur even in the continuous model of spherical dust systems \cite{szekeres}.
Note, however, that they are also frequently referred to as `shell crossing singularities'.
These type of singularities are known to be much weaker than the central ones; nevertheless, in the continuous model the evolution stops whenever they occur.

Returning to the interpretation of panels (a)-(d) of figure \ref{figure4},
and (g)-(h) of figure \ref{figure5}, by comparing these plots it is straightforward to recognize
that either type of deviation from the uniform distribution yields a visible dispersion of the shells.
The focusing of the shells is lost faster and the spreading is more intensive as the initial distributions get further and further away from uniformity.

\begin{figure}
\center
\includegraphics[width=78mm,angle=270]{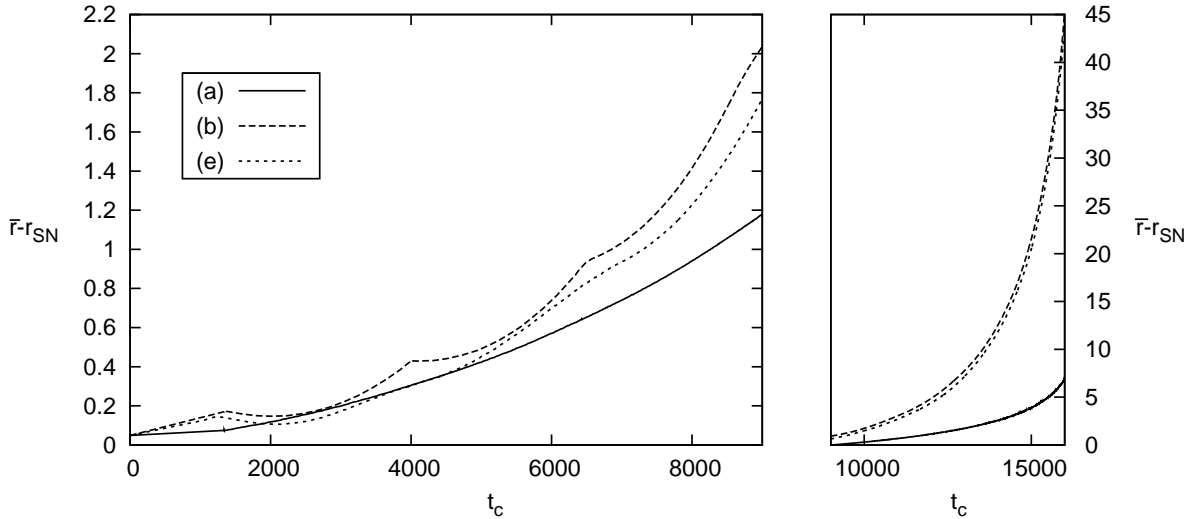}
\caption{The time dependence of the differences of the radial center of mass values of the shell systems depicted in panels (a), (b) and (e)
in figures \ref{figure4} and \ref{figure5} and that of the Newtonian reference solution of a single shell with mass $M_{\rm shell}=16$
are indicated by plotting the $\bar r(\TC)-r_{\rm NS}(\TC)$ functions, where $\TC$ denotes the Minkowski time in the central region.
$\bar r(\TC=0)=2007.5$ and $\bar r-r_{\rm NS}=0$ at $\TC=0$ for each of the individual systems.}
\label{figure6}
\end{figure}

Let us now make some more specific comments. Comparing panels (a) and (b),
depicting the Newtonian and relativistic evolution of completely uniform distributions,
it is visible that the nonlinearity of the relativistic evolution yields a larger dispersion indicated by the fact
that in the Newtonian case the first five knots are pretty well focused,
while the loss of the coherence starts earlier in the relativistic case.
Panels (c) and (d) of figure \ref{figure4} justify that the main characters of the evolution do not change whenever either
the number of the shells is increased or the EOS is changed a little.
In panel (e), where the initial mass distribution is centered, as it is specified by (\ref{mmm}),
knots do not develop and at the beginning the dispersion of the system dominates.
On the other hand, during the second half of the indicated period the shells start to be concentrated around two of the highest mass shells,
and then the groups formed this way start to oscillate around each other.
In panel (f), where instead of the mass distribution the initial radial distribution is concentrated,
a similar pair of groups of shells can be seen to develop. In both panels (e) and (f)
it is also visible that the radius of the outermost shell varies on a significantly larger scale than in the other panels.
In particular, their oscillation amplitudes increase during the initial part and start to decrease only later.
Nevertheless, the relative variety on their motion, with respect to the characteristic size of the shells
composed of all but the outermost shells, remains significant during the entire evolution.
The gray shading makes this type of effect more apparent.
In panels (g) and (h) the initial part of the distribution of the shells is random.
It is clearly visible in these panels that the shells are not concentrated into two groups as before.

In order to make the differences of the above-discussed dynamics more approachable,
the time dependence of the differences, $\bar r-r_{\rm NS}$ of the radial center of mass of the shell system depicted
in panels (a), (b) and (e) in figures \ref{figure4} and \ref{figure5} is shown in figure \ref{figure6},
where $\TC$ denotes the Minkowski time of the central region and $r_{\rm NS}=r_{\rm NS}(\TC)$ stands for
the time dependence of the Newtonian reference solution of a single shell with mass $M_{\rm shell}=16$.
In the left panel of figure \ref{figure6} a zoom into the final part of the evolution is shown in order to make the slight differences more visible.
Note that the wavy figures do correspond to the formation knots indicated by the initial parts of the time evolution
in panels (b) and (e) of figures \ref{figure4} and \ref{figure5}.
It is worth emphasizing that the radial center of mass, $\bar r=\bar r(\TC)$,
is a monotonically decreasing function for each of the individual systems.

\subsection{Dispersion of small perturbations}

\begin{figure}
\center
\vspace{4mm}
\hspace{25mm}\includegraphics[width=45mm,angle=270]{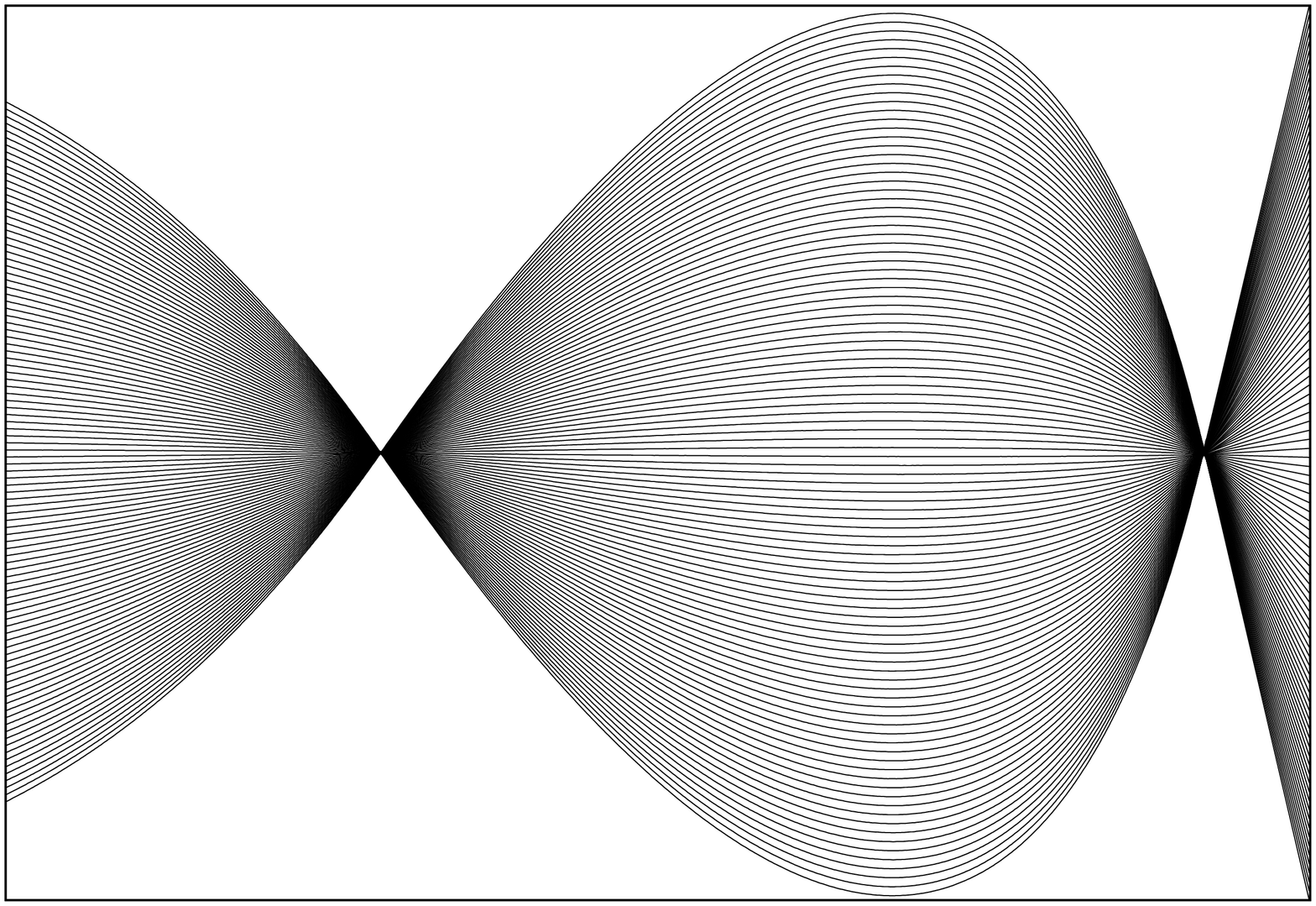}
\includegraphics[width=45mm,angle=270]{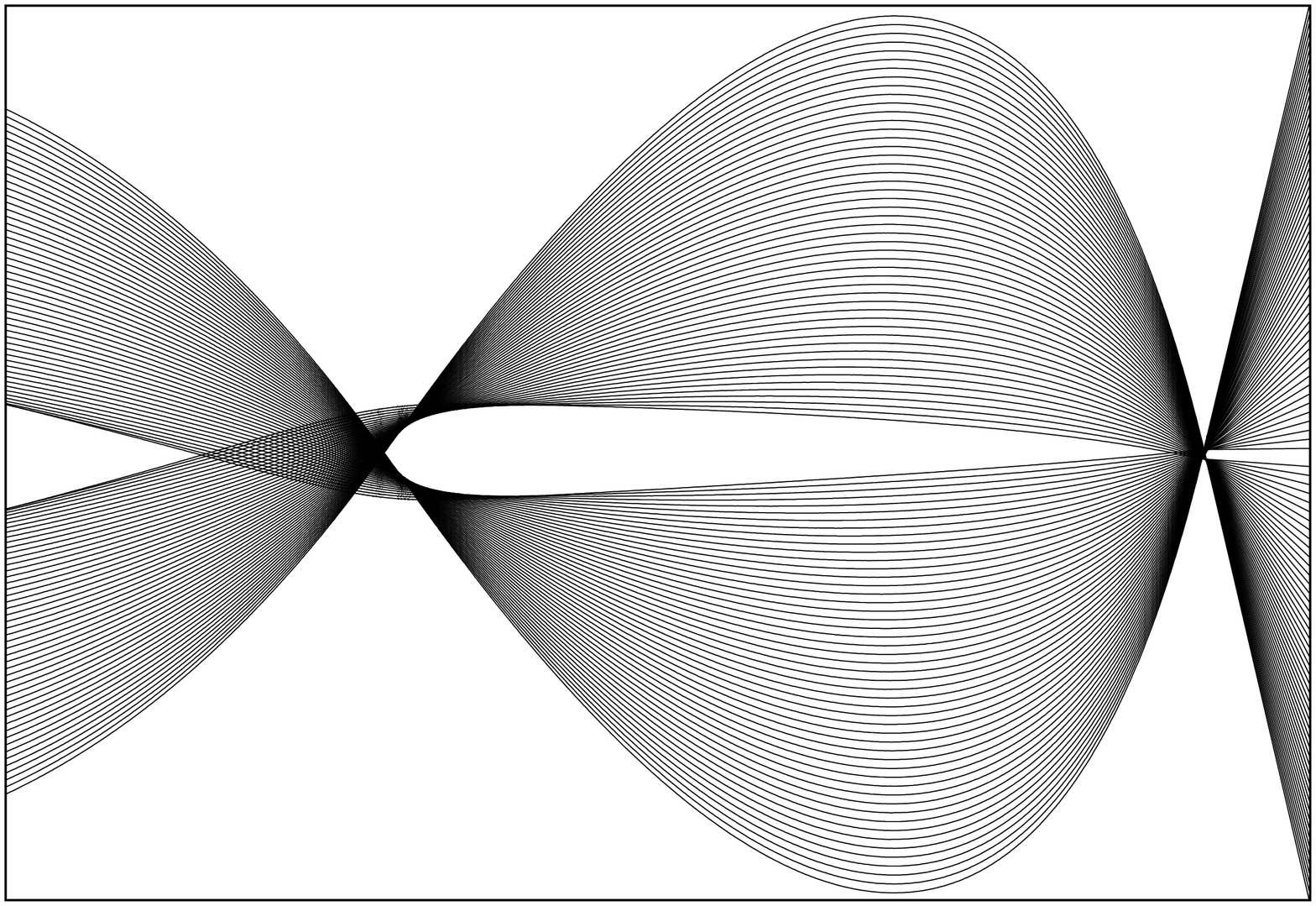}\\
\hspace{25mm}\includegraphics[width=45mm,angle=270]{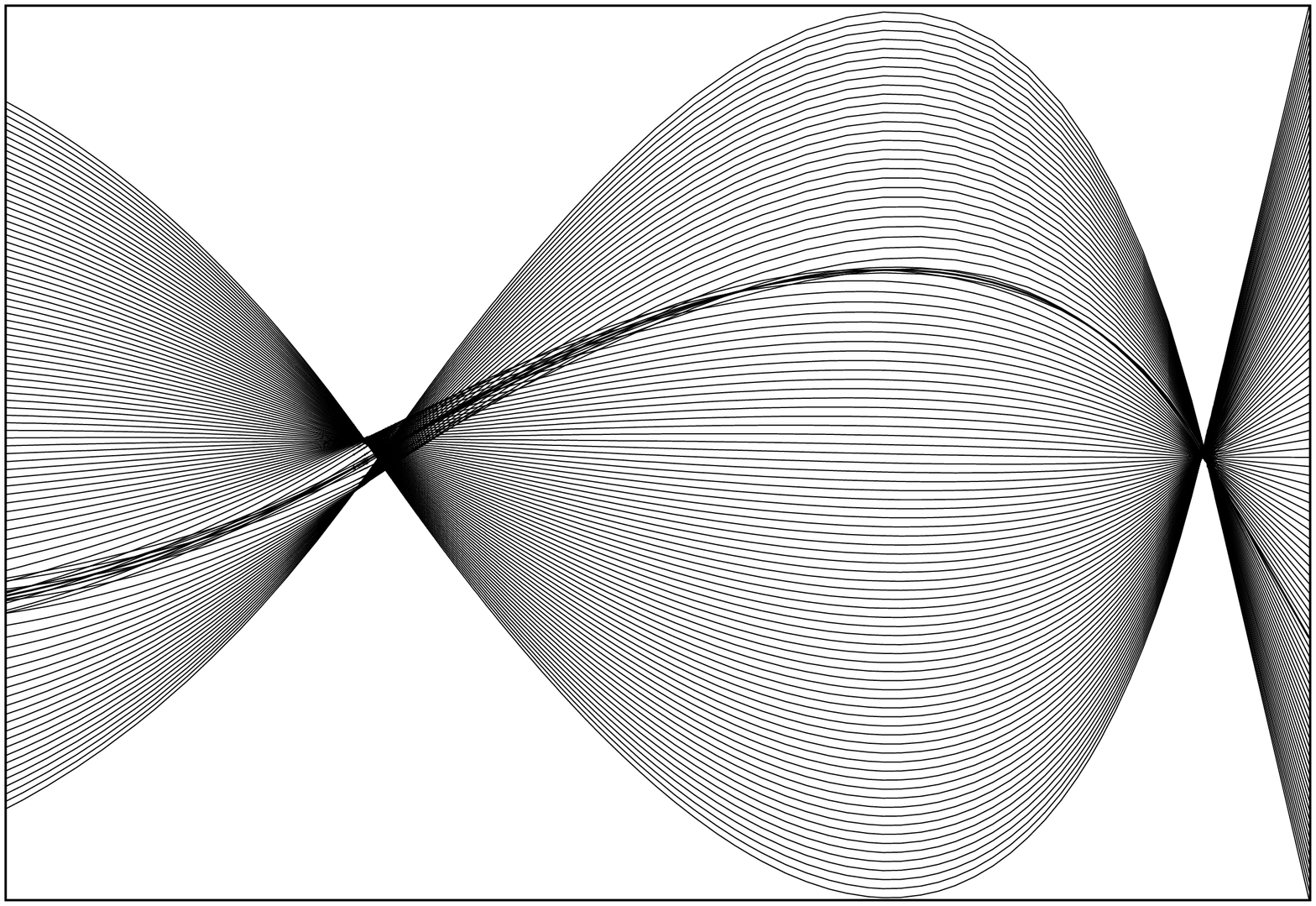}
\includegraphics[width=45mm,angle=270]{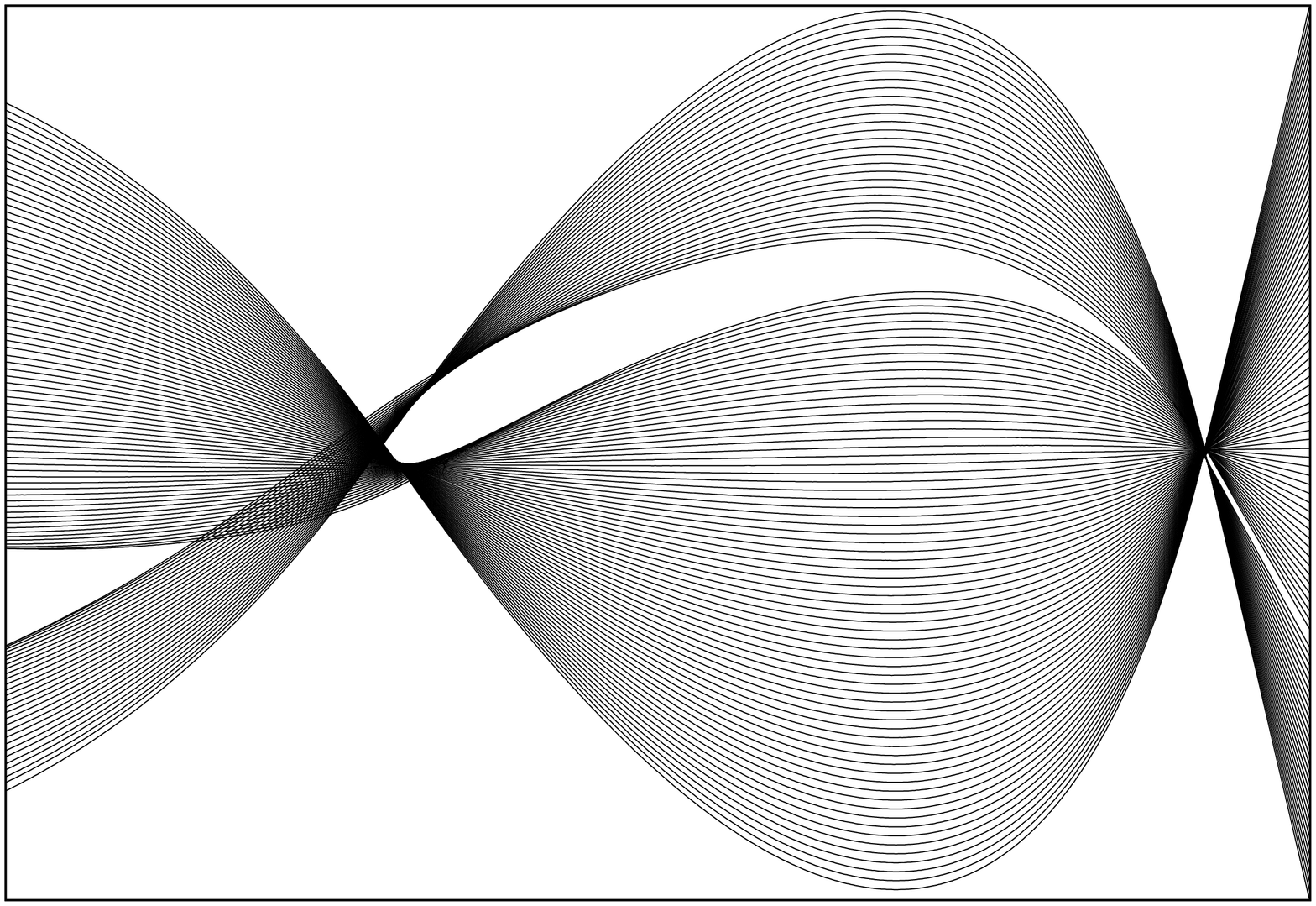}\\
\hspace{25mm}\includegraphics[width=45mm,angle=270]{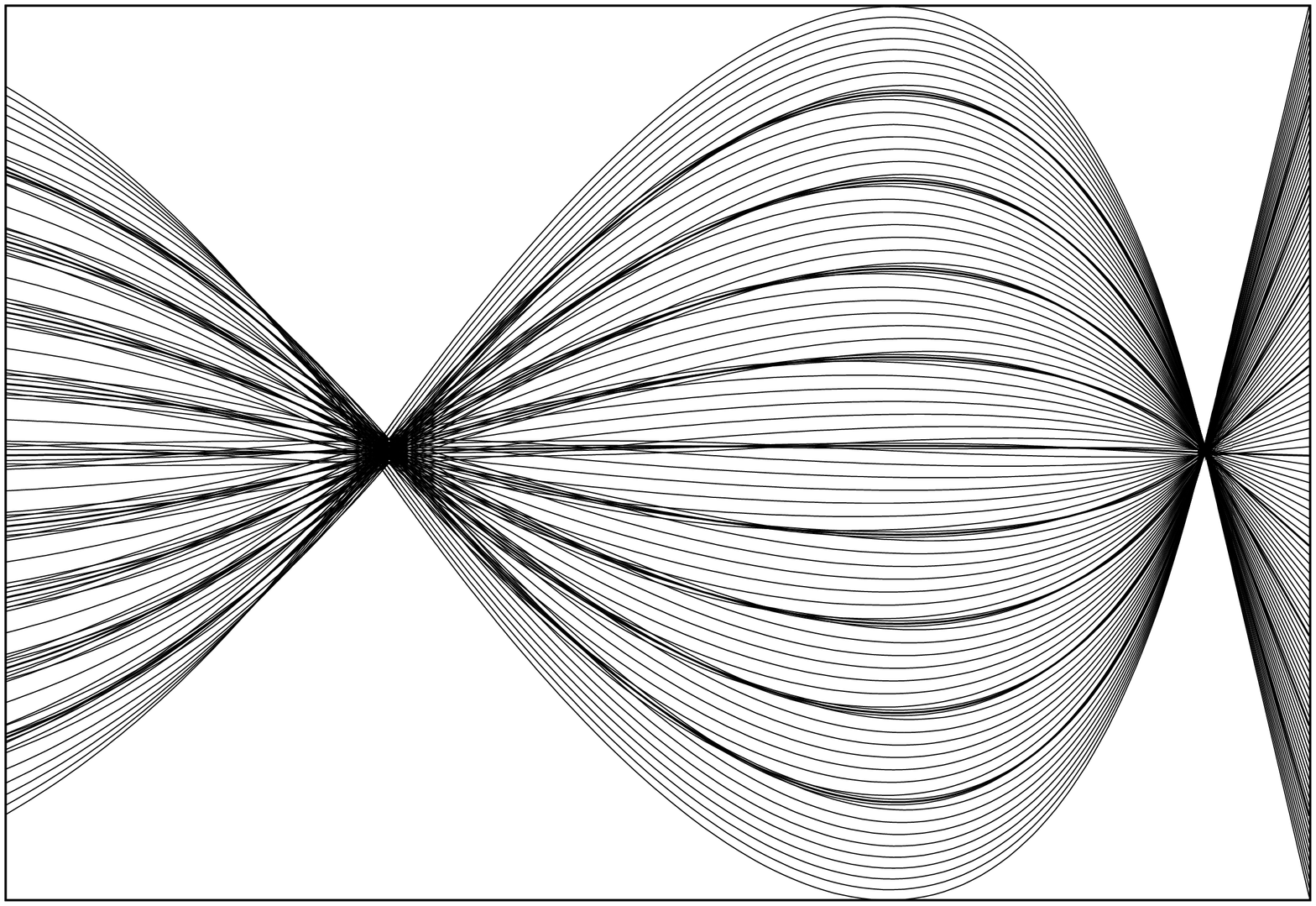}
\includegraphics[width=45mm,angle=270]{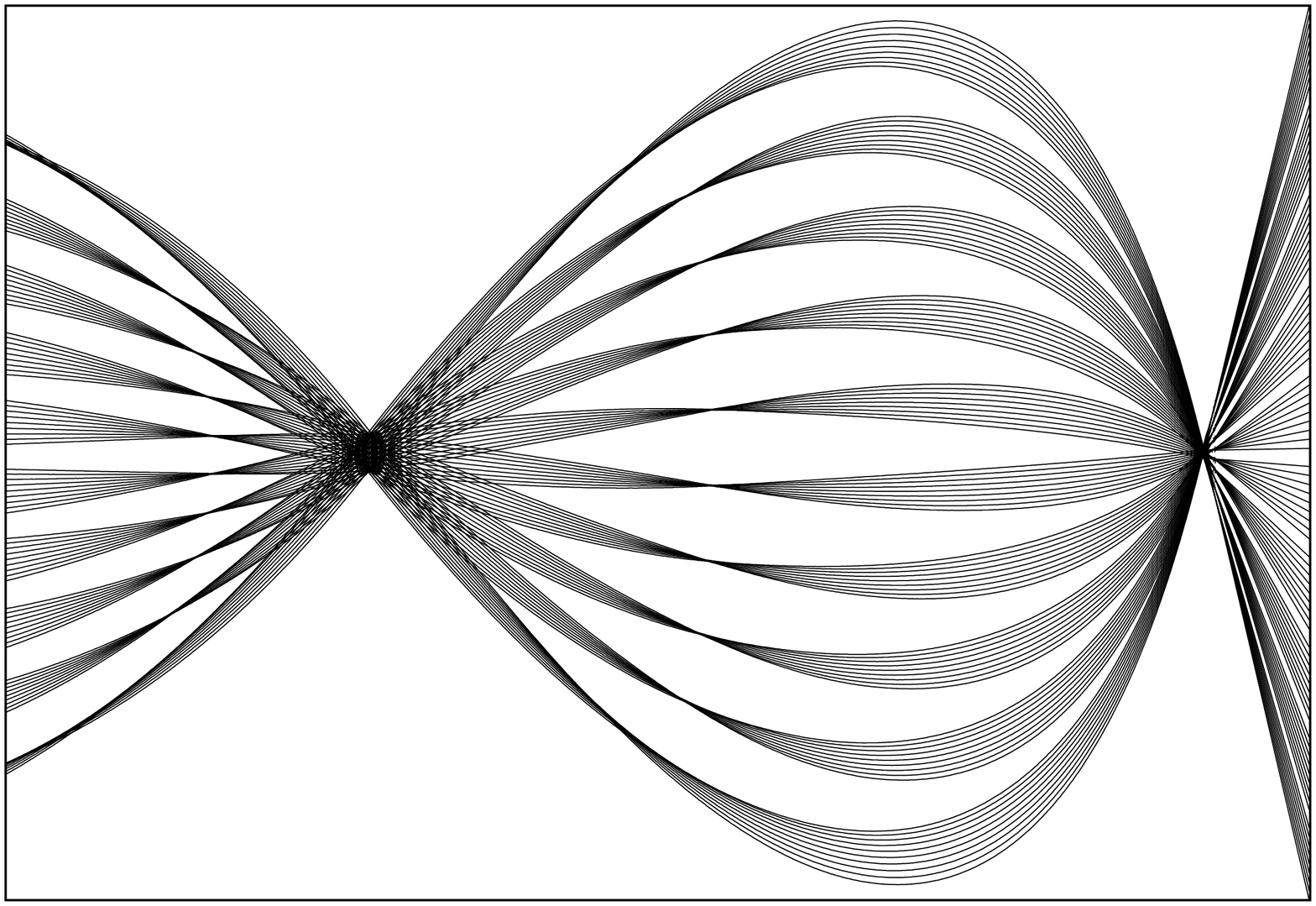}
\vspace{2mm}
\caption{In these plots the initial parts of the evolution of shell systems yielded by slight perturbations of a reference system---consisting of
101 uniformly distributed equal mass shells starting from rest, depicted by the {\it upper-left} panel---are shown.
The initial radial distribution, $r^{(i)}_0$, of the reference system is chosen such that the location of the shells is symmetric to
$r^{(51)}_0$ with $r^{(i)}_0=9999+i$, while ${m}^{(i)}_0=1$ and ${v}^{(i)}_0=0$ for $i=1,2,...,101$.
The systems with `gaps' shown on the {\it upper-right} and {\it middle-right} panels are the result of removing
the 51th and 31th shells from the reference system, respectively.
The {\it middle-left} panel shows a system with an `anti-gap' where the mass of the 31th shell is doubled,
i.e.\,${m}^{(31)}_0=2$. In the {\it lower panels} the evolution of systems with ten uniformly distributed gaps or anti-gaps is shown.
The radial center of mass $\bar r$ is indicated on the horizontal axis---which takes values from the interval $[9425,10050]$
on each panel--- while in the vertical direction the deviations, $\Delta r^{(i)}=r^{(i)}-\bar r$, are shown.}
\label{figure7}
\end{figure}

In this subsection the number of shells is increased significantly from N=16 to 101 in our reference simulation
with uniform initial mass and radial distributions shown by the upper-left panel of figure \ref{figure7}.

\begin{figure}
\center
\includegraphics[width=45mm,angle=270]{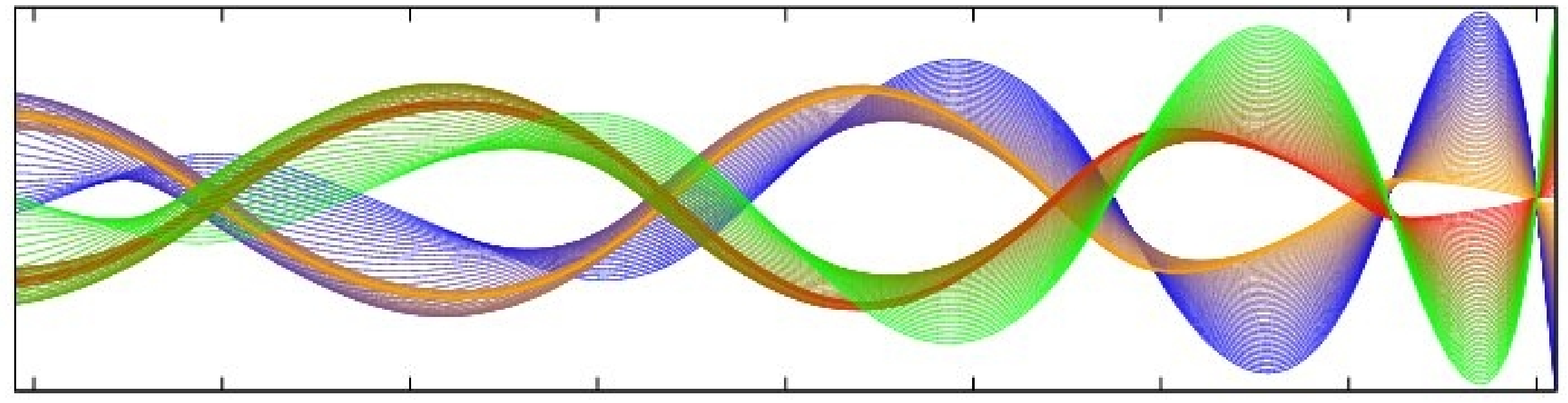}
\caption{In this colored plot a longer period of the time evolution,
for the system in the {\it upper-right} panel of figure \ref{figure7}, is shown.
It demonstrates that a slight change of the uniform initial data might result significant differences in the long run,
and the formation of `crusts' and reversing of sides are also noticeable.}
\label{figure8}
\end{figure}

The applied small perturbations on the evolution of equal mass shell systems
are produced either by doubling the mass of a single shell (or ten shells) of an initially uniform distribution,
or by taking out a single shell (or ten shells) from an initially uniform distribution.
The initial part of the evolutions of these slightly perturbed systems are compared to that of
an initially uniform distribution in figure \ref{figure7}.

The evolutions relevant for systems where the rest mass of one or ten shells is doubled are shown in the middle-left and lower-left panels respectively.
The change of the motion of the other shells is noticeable but at first glance it is not too striking.
As opposed to this, in the panels on the right column of figure \ref{figure7}---where in the upper and middle panels
only a single shell is removed, while ten shells are taken out from the initially uniform distribution in the lower panel---the perturbations
yield more significant changes. The consequences of the indicated small changes are more
significant in spite of the fact that the change of the total rest mass is about 1\% whenever only a single shell is left out
from the reference simulation with uniform initial distributions.

By the inspection of the figure above,
the following important qualitative observations can be made.
First of all, there is a tendency for the formation of a `crust'---represented by
the increase of the density of shells---at the edges of the widening gaps.
Second, following the widening of a gap, a reversing of the sides also occurs,
i.e.,\,the innermost shell becomes the outermost and vice versa, as it is clearly visible in the colored figure \ref{figure8}.
In addition, the crusts are `growing' and during the collapse the entire system gradually becomes a dispersed two-shell system.
Third, around the anti-gaps an increase in the density of shells is also noticeable,
although the growing rate is much lower than in case of gaps.
The two lower panels in figure \ref{figure7} indicate that
the evolution of systems with gaps and anti-gaps get closer to each other when a large number of gaps and anti-gaps
are uniformly distributed in the initial configurations.

\subsection{Mass inflation}
\label{minf}

In starting this subsection we would like to mention that some of the arguments below
were motivated by claims of \cite{Nakao99} about the time evolution of a system formed by a pair of
repeatedly intersecting equal rest mass shells. However, we would like to emphasize that
our conclusions about the possible rate of the blowing up of the mass of the intermediate regions---these are
also justified by means of numerical investigations below---differs from that of \cite{Nakao99}.

\begin{figure}
\center
\includegraphics[width=86mm,angle=270]{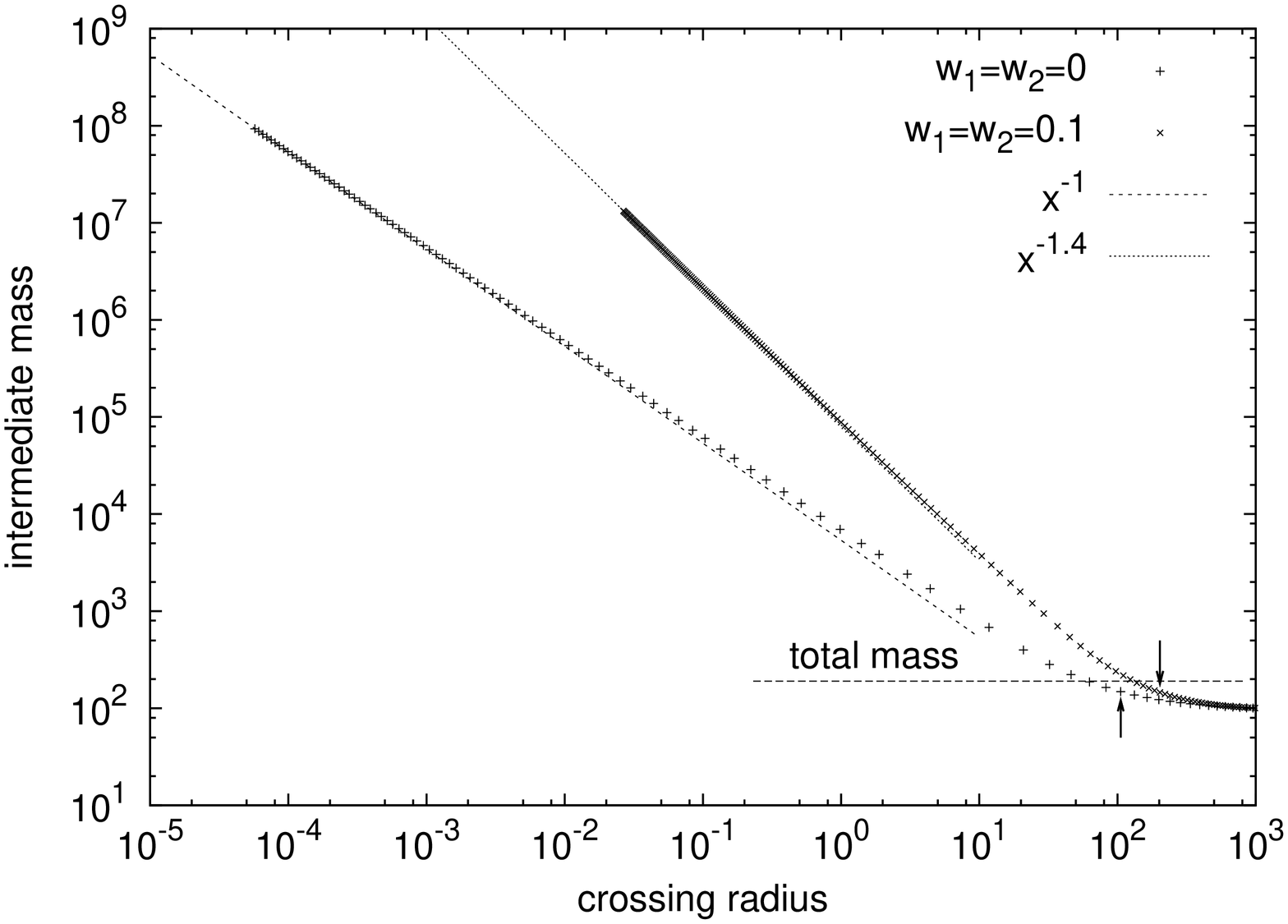}
\vspace{2mm}
\caption{The {\it log-log} plot of the mass parameter of the intermediate region with respect to the radial coordinate
of the collisions is shown for a system formed by two repeatedly intersecting equal rest mass shells possessing linear EOS with $w_1$ and $w_2$.
Note that the dots in this figure signify, at the location of the collision,
the value of the mass of the intermediate region produced in the collision,
while between any two collisions the mass of the intermediate region is constant.
The plot with $w_1=w_2=0$ corresponds to the case of colliding dust shells.
In both cases the initial data were synchronized in the Schwarzschild time of the intermediate region
and was chosen to be the same as in \cite{Eid00}, i.e.\,\,such that
$m_\mathrm{S}=0$, $m_0^{(1)}=100$, $\MG{}^{(1)}=100$, $m_0^{(2)}=100$, $\MG{}^{(2)}=90.05906$.
(Note that in virtue of (\ref{mg}) $\MG{}^{(1)}$ and $\MG{}^{(2)}$ determine the initial values $v_0^{(1)}$ and $v_0^{(2)}$.)
The low radius behavior of the plotted curves justifies that the Schwarzschild mass of the intermediate regions
grows as $\overline M\sim r^{-\alpha}$, where for $\alpha$ the fits to the estimate $\alpha=1+2w_1+2w_2$ formulated by (\ref{transparentp8b}).
The arrows point to the locations where the critical radii of the external region become larger than the radii
of the succeeding collisions, i.e. where $\epsilon_{t_+}$ for the outer shells change their signs.}
\label{figure9}
\end{figure}

In describing the unbounded growth of mass, recall first that the squares of energy and momentum exchanges,
$\Delta \mathcal{E}$ and $\Delta p$---for their definitions see subsection \ref{collshells}---are related as

\begin{equation}
\label{transparentp7}
\Delta \mathcal{E}^2-\Delta p^2=\frac{\MR{}_{1}^2\MR{}_{2}^2}{\RC{}^2}
\end{equation}

\noindent
which implies that for $\Delta \mathcal{E}$ the inequality

\begin{equation}
\label{transparentp7a}
\Delta \mathcal{E}\geq \frac{\MR{}_{1}\MR{}_{2}}{\RC{}}\,
\end{equation}

\noindent
holds. Now, by making use of the relations $\MG{}_2=M_3-M_2$ and $\MG{}_4=M_4-M_1$,
along with the vanishing of $\MC{}_1=M_1$, in virtue of (\ref{transparentp6}), we get

\begin{equation}
\label{transparentp8}
M_4+M_2=M_3+\Delta \mathcal{E}\,.
\end{equation}

\noindent
Hence, in virtue of (\ref{transparentp7a}) and (\ref{transparentp8}), for the radial center of mass,
$\overline M=(M_4+M_2)/2$, of the mass parameters of the successive intermediate regions, the relation

\begin{equation}
\label{transparentp8a}
\overline M \gtrapprox \frac12\left(M_3+\frac{\MR{}_{1}\MR{}_{2}}{\RC{}}\right)\,
\end{equation}

\noindent
has to hold. This latter inequality, whenever $\RC{}$ tends to zero and $\Delta \mathcal{E}$ becomes much larger than $M_3$---for the case
of colliding shells possessing linear EOS with $w_1=c_{\mathrm{s1}}^2$ and $w_2=c_{\mathrm{s2}}^2$---implies that
$\overline M$ must tend to infinity such that the asymptotic blow-up rule

\begin{equation}
\label{transparentp8b}
\overline M \gtrapprox {C}{r^{-(1+2w_1+2w_2)}}\,
\end{equation}

\noindent
holds.

The process, which in repeated collisions of a pair of shells leads to an unbounded growth of the mass parameter of the intermediate region,
is the mass inflation. The blow-up behavior of some simple configurations, along with the justification of the estimate (\ref{transparentp8b}),
is shown in figure \ref{figure9}.

Some remarks are now in order.
In explaining the relatively small values of $w$ it should be mentioned that
whenever larger values of $w_1$ and $w_2$ are applied, or $w_1$ and $w_2$ differ significantly,
the fluid shells start to move outwards and it may happen that they collide only once or they do not collide at all.
It is also important to note that whenever the collapse and mass inflation occur the blow-up rate behavior is insensitive
to the initial data of the shells.

\begin{figure}
\center
\includegraphics[width=86mm,angle=270]{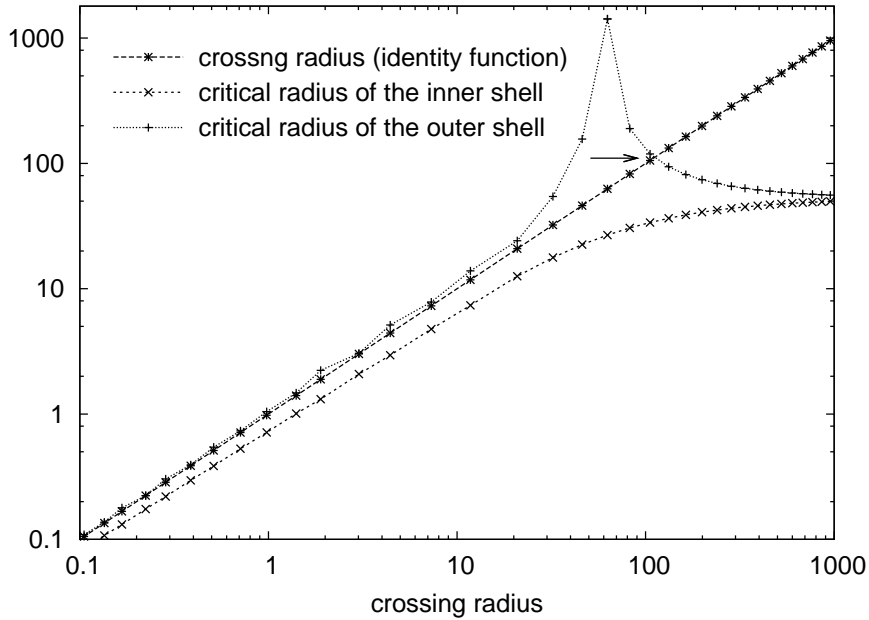}
\vspace{2mm}
\caption{The values of $\widehat r_{\rm inner}$ and $\widehat r_{\rm outer}$ as a function of $\RC{}$
are plotted on a {\it log-log} scale for the case of repeatedly colliding equal rest mass dust shells corresponding to the `points'
represented by the `$+$' sign in figure \ref{figure9}. The intersection of the $\RC{}=\widehat r_{\rm outer}(\RC{})$
curves signifies---indicated by the arrow---the location where $\epsilon_{t_+}$ of the outer shell changes sign.
Note also that gravitational mass of the outer shell changes sign where $\widehat r_{\rm outer}$ attains its maximum.}
\label{figure10}
\end{figure}

Note that mass inflation is not new; it is known to occur in the continuum limit (see, e.g, \cite{csizmadia} for a recent numerical investigation).
Nevertheless, at first sight the occurrence of the mass inflation in the thin shell formalism is unexpected because it is known
that the mass of the intermediate region cannot be larger than that of the outer region unless the radius of the collision,
$\RC{}$, becomes smaller than the critical radius, $\widehat r$, of the outer shell.
Note that such a critical value in the case of the dynamics of a single shell is extremely small,
much smaller than the Schwarzschild radius. In order to show that the above-formulated expectation is justified
by the investigated time evolutions in figure \ref{figure10}, the time dependence of the critical radii of the inner and outer shells---their
relative location varies in time---along with the time dependence of the radius of the collision is plotted.
The location where for the first time $\RC{}$ becomes smaller than $\widehat r$ for the temporarily outer shell
is the very location where the mass of the intermediate region, $M_4$, becomes larger than the mass of the outer spacetime, $M_3$.

\begin{figure}
\center
\vspace{2mm}
\includegraphics[width=142mm,angle=0]{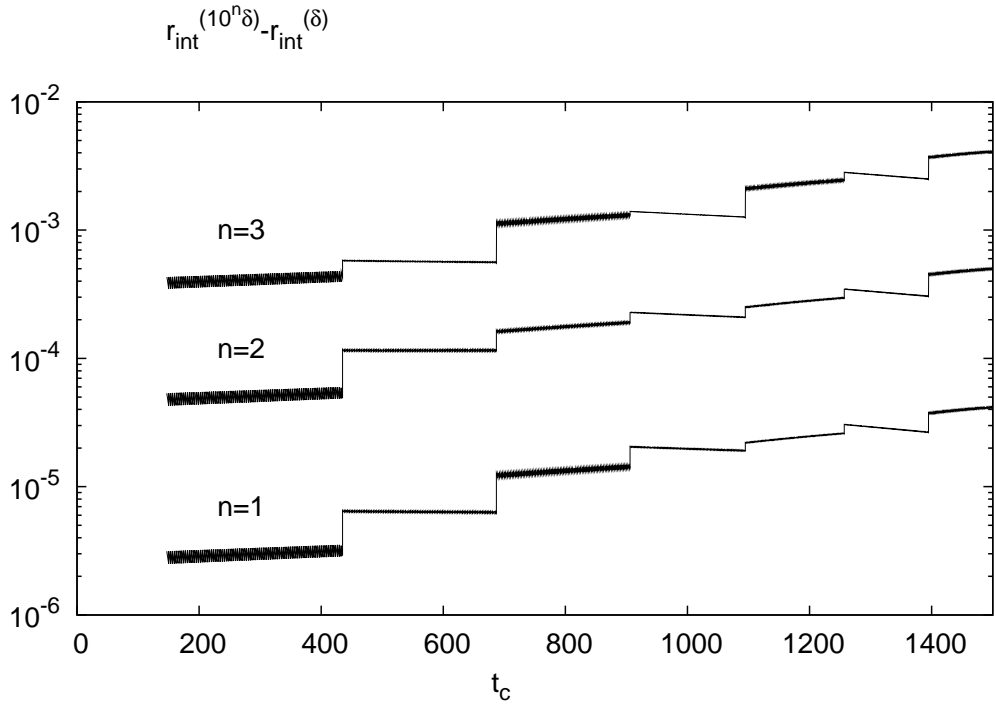}
\caption{The time dependence of the difference, $r_{\rm int}^{(10^{n}\delta)}-r_{\rm int}^{(\delta)}$,
of the numerical value of the radius of the (temporarily) internal shell is plotted relevant to resolutions
$10^{n}\delta$ with $n=1,2,3$ and for the reference numerical solution $r_{\rm int}^{(\delta)}$.}
\label{figure11}
\end{figure}

The precision of the applied numerical schema is determined in the case of the above-described system
formed by two repeatedly intersecting equal rest mass dust shells.
Denote by $r_{\rm int}^{(\Delta)}$ the numerical value of the radius of the (temporarily) internal shell
relevant for resolution $\Delta$. In figure \ref{figure11} the time dependence of the difference $r_{\rm int}^{(10^n\delta)}-r_{\rm int}^{(\delta)}$
is plotted for the initial period for $\Delta=10^n\delta$ with $n=1,2,3$ and $r_{\rm int}^{(\delta)}$ denotes
the reference numerical solution with the smallest resolution $\delta$.
As is expected for any fixed value of $n$, the difference $r_{\rm int}^{(10^n\delta)}-r_{\rm int}^{(\delta)}$
is increasing in time, however, by decreasing the value of $n$ by 1 yields an order of magnitude downward shift
of the successive curves. This justifies that our numerical code is convergent even though collisions---indicated by the jumps on the curves---occur.

Let us finally mention that mass inflation is not specific only to two-shell systems as it did occur for systems consisting of more than two shells.
According to our experience the masses of each of the intermediate regions increase apparently following the above-derived power law rule,
but there was a definite order kept during the evolution. More definitely, at any instant of the Eddington--Finkelstein time
the mass of an inner intermediate region was, in all of our simulations, smaller than the mass of any of the intermediate regions
located outwards with respect to the inner one.

\section{Final remarks}
\setcounter{equation}{0}

The relativistic time evolution of multi-layer spherically symmetric shell systems has been investigated.
After recalling the basics of the analytic setup a newly developed numerical code is introduced.
This numerical method was made to be capable of following the time evolution of systems comprising of great numbers of colliding shells
such that whenever collisions occur they are assumed to be totally transparent.

By making use of our numerical method for the first time, the relativistic time evolution of numerous shell systems
involving large number of thin shells could be made. As these systems may be considered as approximate models of thick shells,
the results reported in this paper provide insights about their dynamical behavior as well.
The most important observations we have made can be characterized by the key phrases:
concentrations of subsets of shells, formation of `crusts' at the boundaries, reversing of the sides of gaps.

The chosen analytic setup ensured that the evolution of the considered shell systems can be investigated
both in the domain of outer communication and in the black hole region.
This made our numerical code capable of studying mass inflation within the thin shell formalism.
We would like to emphasize that beside the numerical investigation of mass inflation,
an estimate explaining the main features of the blow-up behavior of the mass parameter of the intermediate region is also provided.

As was mentioned in the introduction, there are a great number of astrophysical systems which can be modeled by shells.
For instance, there are plans to investigate the interaction of repeated quasi-spherical matter ejections by supernovas
using the developed method.

\section*{Acknowledgments}

This research was supported in part by OTKA grant K67942.

\end{document}